\documentclass[
 reprint,onecolumn,
 amsmath,amssymb,
 aps,nofootinbib,superscriptaddress,
]{revtex4-2}

\usepackage{graphicx}
\usepackage{dcolumn}
\usepackage{bm}

\newcommand{\SupplSectionTIPC}{S1~}
\newcommand{\SupplSectionESP}{S2~}
\newcommand{\SupplSectionESN}{S3~}
\newcommand{\SupplSectionAttractorAnalysis}{S4~}

\newcommand{\SupplSectionTIT}{S6~}

\newcommand{\ba}{\boldsymbol{a}}

\newcommand{\bb}{\boldsymbol{b}}
\newcommand{\bc}{\boldsymbol{c}}

\newcommand{\boldf}{\boldsymbol{f}}
\newcommand{\boldF}{\boldsymbol{F}}
\newcommand{\bg}{\boldsymbol{g}}
\newcommand{\bh}{\boldsymbol{h}}
\newcommand{\bI}{\boldsymbol{I}}
\newcommand{\bJ}{\boldsymbol{J}}
\newcommand{\bQ}{\boldsymbol{Q}}
\newcommand{\bR}{\boldsymbol{R}}
\newcommand{\bu}{\boldsymbol{u}}
\newcommand{\bx}{\boldsymbol{x}}
\newcommand{\bX}{\boldsymbol{X}}
\newcommand{\by}{\boldsymbol{y}}
\newcommand{\bz}{\boldsymbol{z}}
\newcommand{\bw}{\boldsymbol{w}}
\newcommand{\bW}{\boldsymbol{W}}

\newcommand{\uth}{^{\rm th}} 
\newcommand{\xt}{{\bm x}_t}
\newcommand{\Ctot}{C_{\rm tot}}

\newcommand{\bgamma}{\boldsymbol{\gamma}}

\newcommand{\br}{\boldsymbol{r}}

\newcommand{\bSigma}{\boldsymbol{\Sigma}}
\newcommand{\bU}{\boldsymbol{U}}

\newcommand{\bV}{\boldsymbol{V}}
\newcommand{\bY}{\boldsymbol{Y}}
\newcommand{\bZ}{\boldsymbol{Z}}

\begin{document}

\title{
    Reservoir Computing Generalized
}

\author{Tomoyuki Kubota}
 \email{kubota@isi.imi.i.u-tokyo.ac.jp}
 \affiliation{Graduate School of Information Science and Technology, The University of Tokyo, Japan}
 \affiliation{Next Generation Artificial Intelligence Research Center (AI Center), The University of Tokyo, Japan}
\author{Yusuke Imai}
 \affiliation{Graduate School of Information Science and Technology, The University of Tokyo, Japan}
\author{Sumito Tsunegi}
 \affiliation{National Institute of Advanced Industrial Science and Technology, Japan}
\author{Kohei Nakajima}%
 \affiliation{Graduate School of Information Science and Technology, The University of Tokyo, Japan}
 \affiliation{Next Generation Artificial Intelligence Research Center (AI Center), The University of Tokyo, Japan}

\date{\today}

\begin{abstract}
A physical neural network (PNN)~\cite{lee2022mechanical,ross2023multilayer,xue2024fully} has both the strong potential to solve machine learning tasks and intrinsic physical properties, such as high-speed computation and energy efficiency. 
Reservoir computing (RC)~\cite{jaeger2001echo,maass2002real,nakajima2021reservoir} is an excellent framework for implementing an information processing system with a dynamical system by attaching a trained readout, 
thus accelerating the wide use of unconventional materials for a    PNN~\cite{kaspar2021rise,yasuda2021mechanical,cazettes2023reservoir}. 
However, RC requires the dynamics to reproducibly respond to input sequence~\cite{jaeger2001echo}, 
which limits the type of substance available for building information processors. 
Here we propose a novel framework called generalized reservoir computing (GRC) by turning this requirement on its head, 
making conventional RC a special case. 
Using substances that do not respond the same to identical inputs
(e.g., a real spin-torque oscillator), we propose mechanisms aimed at obtaining a reliable output and show that processed inputs in the unconventional substance are retrievable. 
Finally, we demonstrate that, based on our framework, spatiotemporal chaos, which is thought to be unusable as a computational resource, can be used to emulate complex nonlinear dynamics, including large scale spatiotemporal chaos. 
Overall, our framework removes the limitation to building an information processing device and opens a path to constructing a computational system using a wider variety of physical dynamics. 
\end{abstract}

\maketitle

The physical instantiation of neural networks is an urgent challenge to solve issues caused by traditional computers, such as energy consumption for computation~\cite{lecun2015deep,merolla2014million,prezioso2015training}. 
Unconventional materials have intrinsic physical properties that are not found in the traditional substance of semiconductors, such as low energy consumption~\cite{ross2023multilayer,xue2024fully} and high-speed processing~\cite{xue2024fully}, potentially allowing us to build a computational system with innovative physical properties. 
From the perspective of information processing, a key factor is the dynamical state in the physical system, which works as memory and nonlinear processing units. 
One of the most remarkable frameworks to physicalize neural networks is physical reservoir computing (PRC)~\cite{nakajima2020physical}.

Reservoir computing (RC)~\cite{jaeger2001echo,maass2002real,nakajima2021reservoir} is a machine learning framework for implementing an information processing system using a dynamical system, which provides the theoretical basis for PRC. 
In the RC framework, we inject inputs into the system and obtain outputs by attaching a trained readout, which can be either a (frequently used) linear function~\cite{jaeger2001echo} or a nonlinear one~\cite{maass2002real}. 
This simple configuration is applicable to any dynamical system, including natural physical dynamics, to realize various types of physical reservoirs (e.g., electronics~\cite{appeltant2011information,du2017reservoir,zhong2022memristor}, quantum systems~\cite{fujii2017harnessing,ghosh2021quantum,mujal2021opportunities,kubota2023temporal}, optical components~\cite{brunner2013parallel,larger2012photonic}, spintronics~\cite{torrejon2017neuromorphic,tsunegi2019physical,tsunegi2023information}, mechanical structures~\cite{nakajima2014exploiting,nakajima2015information,akashi2024embedding}, and organisms~\cite{cai2023brain,sumi2023biological,ushio2023computational}). 
However, RC imposes a condition of echo state property (ESP)~\cite{jaeger2001echo}, in which the responses of the system be a function of only the previous input sequence (See Materials and Methods for further details). 
We call this a time-invariant (TI) state~\cite{boyd1985fading,maass2002real} throughout this paper, as it guarantees a reproducible response to the same input sequence. 
Using the TI state and the trained readout, we can realize a required input-output relation in the entire computational system. 
Conversely, this condition limits the range of computational resources to the TI state, thereby excluding a great variety of oscillatory or chaotic materials included in the time-variant (TV) state.

\begin{figure*}[thb]
    \centering
    \includegraphics[scale=0.35]{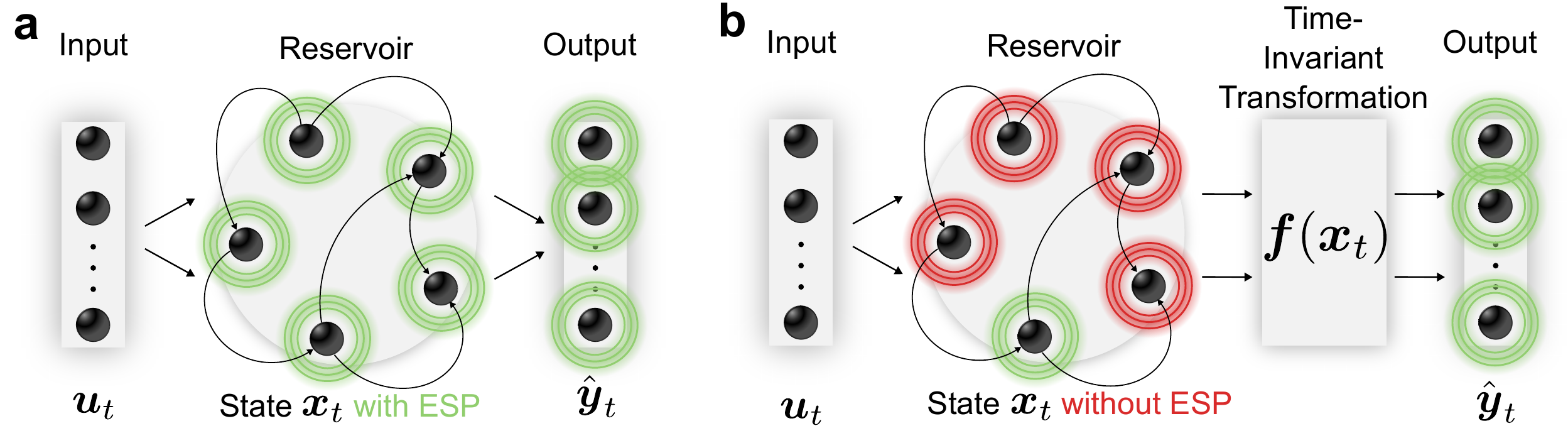}
    \caption{
        \textbf{Conventional RC is a special case of GRC. 
        } 
        \textbf{a}, Conventional and \textbf{b}, GRC frameworks. 
        The reservoir receives the input $\bu_t$ and updates the state $\bx_t$. 
        The outputs $\hat{\by}_t$ are calculated by \textbf{a}, a linear or nonlinear readout and \textbf{b}, a TI transformation, which would be realized through a nonlinear readout with memory $\boldf(\bx_t)$. 
        The green (red) ripples around the nodes represent the state with (without) node-wise ESP (See Supplementary Information~\SupplSectionESP for further details). 
        The conventional RC forms outputs with ESP using the reservoir states, thereby already fulfilling ESP, while the GRC can use not only states with ESP but also those without ESP, whose time-dependence is removed through TI transformation to make outputs equipped with ESP. 
    }
    \label{fig:framework}
\end{figure*}

In this paper, we propose generalized reservoir computing (GRC), a novel framework for exploiting systems, either with or without ESP, as a reliable computational resource. 
To create an information processor with TI outputs, the conventional RC adopts a TI dynamical state as well as a linear or nonlinear readout (Fig.~\ref{fig:framework}a)~\cite{jaeger2001echo,maass2002real}. 
However, reliable information processing requires that the ESP be in the output layer, not in the reservoir layer. 
GRC generates an output with ESP from a general dynamical system to introduce a time-invariant (TI) transformation, regardless of the ESP of the dynamical state in the reservoir (Fig.~\ref{fig:framework}b).

\section*{TI Transformation}
We consider an input-driven dynamical system described by $\bx_{t+1} = \bg(\bx_t,\bu_t)$ with the $N$-dimensional state $\bx_t$ and the $M$-dimensional input $\bu_t$. 
Its solution $\bx_t$ can be regarded as a function of time $t$ and input history $\{\bu_{t-1},\bu_{t-2},\ldots\}$ with an initial state $\bx_0$, $\bx_t=\bh(t,\bu_{t-1},\bu_{t-2},\ldots;\bx_0)$~\cite{kubota2021unifying}. 
The conventional RC imposes the ESP on the reservoir state, which is TI. 
Conversely, if all the states are functions of time and input history called TV 
[i.e., $x_{i,t}=h_i(t,\bu_{t-1},\bu_{t-2},\ldots; \bx_0)$ $(i=1,\ldots,N)$], 
the conventional linear readout cannot always generate an output with ESP. 
The TV state can be transformed into TI outputs by nonlinearity with memory $\boldf(\bx_t,\bx_{t-1},\ldots)$ in general (Note that linear weights can also be used for TI transformation in special cases [See Supplementary Information~\SupplSectionESP for further details]). 
Unless otherwise noted, we focus on TI transformation without memory $\boldf(\bx_t)$ for simplicity from now on (See Supplementary Information~\SupplSectionTIT for examples of TI transformation with memory). 
We call this transformation the TI transformation and define the GRC as follows: 
\begin{align*}
    \bx_{t+1} = \bg(\bx_t, \bu_{t}), ~ 
    \hat{\by}_t = \boldf(\bx_t), 
\end{align*}
where the reservoir state $\bx_t$ is either TI or TV, while the output $\hat{\by}_t$ is TI.

\begin{figure*}[tb]
    \centering
    \includegraphics[scale=0.35]{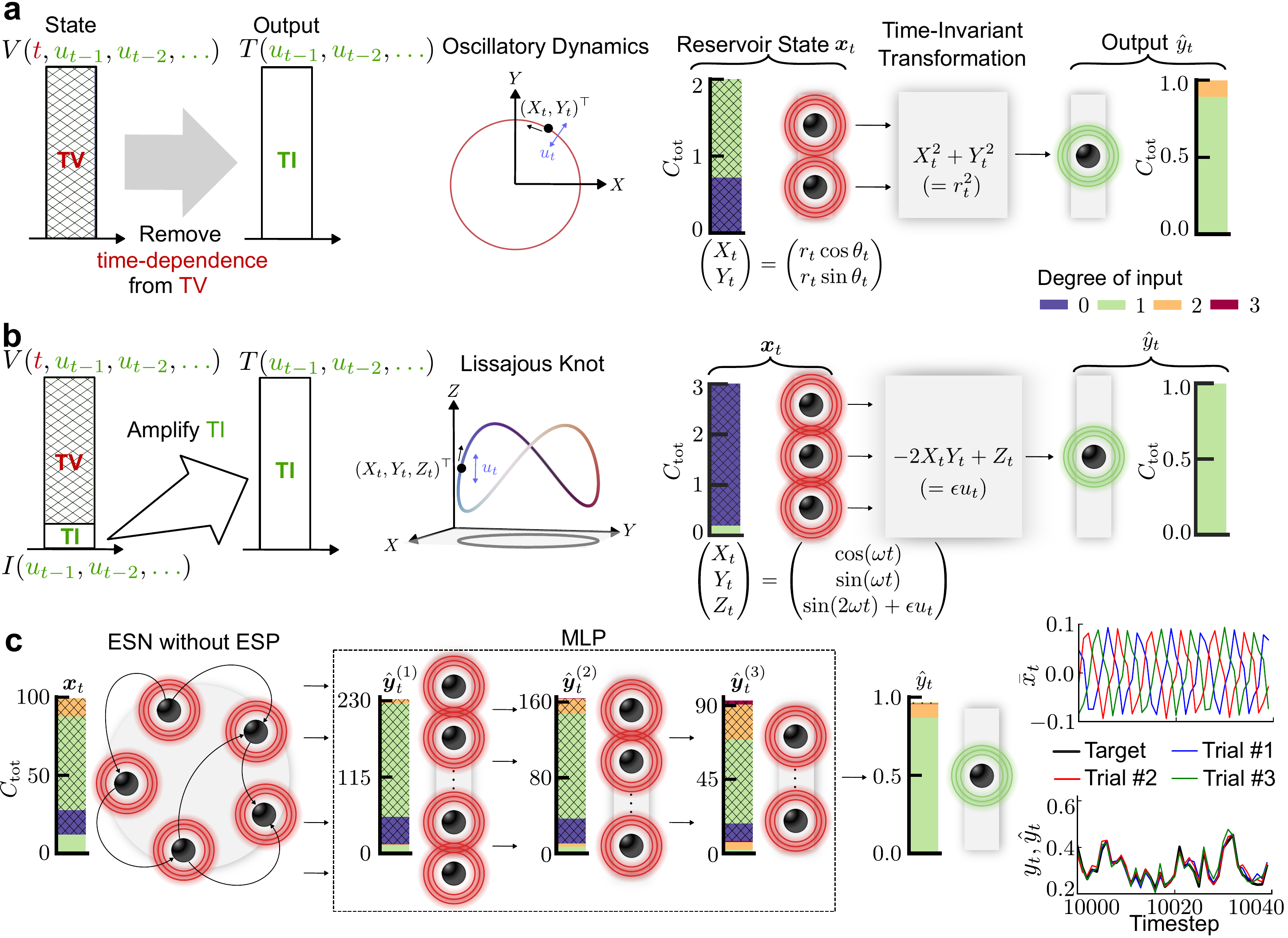}
    \caption{
        \textbf{TI transformation of time-variant reservoirs.}
        \textbf{a}, \textbf{b}, Two analytical examples and \textbf{c}, a numerical transformation, are illustrated. 
        \textbf{a} left, Removal of time-dependence from time-variant (TV, hatched bar) terms to form time-invariant (TI, non-hatched bar) terms. 
        \textbf{a} middle, The trajectory of oscillatory dynamics, on which the state $(X_t, Y_t)^\top$ moves along a circular orbit with a fixed angular velocity and is perturbed by input $u_t$ in the radial direction. 
        \textbf{a} right, Time-dependent elements in the state are canceled out with coordinate transformation to form the TI output. 
        The TIPC decomposition $\Ctot$ is depicted by color bars where the non-hatched and hatched bars represent the TI and TV capacities, respectively. 
        The color represents the degree of input: $0$ (purple), $1$ (green), $2$ (orange), and $3$ (red). 
        \textbf{b} left, Amplification of TI terms by removal of TV terms. 
        \textbf{b} middle, The trajectory of the Lissajous knot, on which the state $(X_t,Y_t,Z_t)^\top$ shows the periodic orbit perturbed by the input $u_t$ in the $Z$-direction. 
        \textbf{b} right, The nonlinear readout enlarges the small TI term in the state by canceling out time-dependent functions (hatched purple). 
        See Fig.~S1 for further details of the TIPC decompositions in \textbf{a} and \textbf{b}. 
        \textbf{c}, The numerical transformation using the ESN with TV states (its maximum conditional Lyapunov exponent was $\lambda_{\rm max}=1.5\times10^{-2}$) and 4-layer MLP. 
        The ripple color represents whether the node holds the ESP (all the node-wise ESP indices $\bar{d}_i<0.3$, green) or not (red). 
        The bar graphs represent the amount of TI and TV terms in the ESN and MLP layers. 
        \textbf{c} right, Time series of the mean-field states $\bar{x}_t$ and outputs $\hat{y}_t$ with three different initial values (trial \#1, blue; \#2, red; \#3, green) and target $y_t$ (black). 
        The normalized mean square errors between $y_t$ and $\hat{y}_t$ were $0.066$ (\#1), $0.068$ (\#2), and $0.071$ (\#3). 
        Note that those with linear regression were $0.40$ (\#1), $0.40$ (\#2), and $0.41$ (\#3). 
    }
    \label{fig:TI_transformation}
\end{figure*}

We introduce two analytical examples of the transformation mechanism. 
The first example removes the time dependence from the TV term $V(t,u_{t-1},u_{t-2},\ldots; \bx_0)$ to form a TI term $T(u_{t-1},u_{t-2},\ldots)$ (Fig.~\ref{fig:TI_transformation}a, left). 
This transformation can be analytically shown using periodic dynamics as an example (Fig.~\ref{fig:TI_transformation}a, middle), which is represented by the radius $r_t$ and the angle $\theta_t$. 
The analytical solution of the radius $r_t$ is the TI function $r(u_{t-1},u_{t-2},\ldots)$, and the angle is the linear function of time $\theta_t=\omega t$. 
The position on the orthogonal coordinate $\bx_t = (X_t, Y_t)^\top = (r_t\cos\theta_t, r_t\sin\theta_t)^\top$ depends both on time $t$ and on input history $\bx_t = (r(u_{t-1},u_{t-2},\ldots) \cos(\omega t), r(u_{t-1},u_{t-2},\ldots) \sin(\omega t))^\top$, which is TV. 
Note that we have taken $(r_0,\theta_0)=(0,0)$ and, from now on, we omit the dependence on the initial value from the following equations for simplicity. 
The TI and TV representations can be revealed by the temporal information processing capacity (TIPC)~\cite{kubota2021unifying,kubota2023temporal}, 
which expands a function $\boldF_t$ with TI function bases $I_i$ and TV bases $V_i$ as: 
\begin{align*}
    \boldF_t = \sum_i \ba_i I_i(u_{t-1},u_{t-2},\ldots) + \sum_i \bb_i V_i(t,u_{t-1},u_{t-2},\ldots)
\end{align*}
and evaluates the amount of processed inputs by the magnitude of coefficients $C_i^{\rm TI}=||\ba_i||^2$ and $C_i^{\rm TV}=||\bb_i||^2$. 
The sums of the TI (TV) capacities are illustrated by a bar graph without (with) hatchmarks for each order of input with different colors. 
For example, the state of periodic dynamics is expanded as $\bx_t=\sum_{\tau}\left(\bc_{\tau} u_{t-\tau}\cos(\omega t) + \bc_{\tau} u_{t-\tau}\sin(\omega t)\right)$, which has only TV terms with zeroth- and first-order input (hatched purple and green in Fig.~\ref{fig:TI_transformation}a, right). 
The conventional RC cannot form an output with ESP from these two states because time-dependent elements cannot usually be canceled out by a weighted sum of the states. 
To remove the time dependence, nonlinear transformation is required in the readout layer in this case. 
This state can be transformed into a TI state [e.g., $f(\bx_t)=X_t^2+Y_t^2=r_t^2(\cos^2\theta_t+\sin^2\theta_t)=r_t^2$, where $r_t^2$ is TI (non-hatched green and orange in Fig.~\ref{fig:TI_transformation}a, right)].

The second example is to amplify the existing TI terms by removing TV terms. 
If the state is represented by a sum of TI and TV terms $I(u_{t-1}, u_{t-2}, \ldots) + V(t, u_{t-1},u_{t-2},\ldots)$, this creates the possibility of transforming the state into TI representation $T(u_{t-1},u_{t-2},\ldots)$ (Fig.~\ref{fig:TI_transformation}b, left). 
We demonstrate the amplification using the Lissajous knot system 
(Fig.~\ref{fig:TI_transformation}b, middle). 
All three states $\bx_t=(X_t,Y_t,Z_t)^\top=(\cos(\omega t), \sin(\omega t), \sin(2\omega t) + \epsilon u_t)^\top$ include time-dependent terms, which make it impossible to form a TI output with linear readout, 
while a nonlinear readout [e.g., $f(\bx_t)=-2X_t Y_t + Z_t=\epsilon u_t$] enables us to calculate an output with ESP (Fig.~\ref{fig:TI_transformation}b, right).

Finally, we introduce a numerical transformation with a nonlinear data-fitting technique 
to cover TI transformation that cannot be obtained in an explicit form, which includes majority of physical systems. 
To find the transformation, we adopted a multi-layer perceptron (MLP) with backpropagation for nonlinear readout. 
We utilized the echo state network (ESN)~\cite{jaeger2001echo} with TV states to perform the NARMA10 benchmark task~\cite{atiya2000new}, whose required memories to solve are known~\cite{kubota2021unifying}. 
Figure~\ref{fig:TI_transformation}c depicts the network structure of the ESN with a 4-layer MLP and the TIPC decomposition for each layer. 
The ESN held both TI and TV inputs, which were converted via MLP layers to the TI output with linear and quadratic past inputs. 
The amounts of required terms are gradually tuned as an increase in layer id (See Supplementary Information~\SupplSectionESN for further details). 
Furthermore, we solved the benchmark task with another ESN time series with different initial values. 
Although the three time series averaged over nodes showed totally different behaviors (Fig.~\ref{fig:TI_transformation}c, right), both outputs $\hat{y}_t$ emulated the target $y_t$ well, indicating success in searching for the numerical transformation. 
These results suggest the possibility of using analytical or numerical transformation to generate the TI output from systems without ESP.

\begin{figure*}[tb]
    \centering
    \includegraphics[scale=0.45]{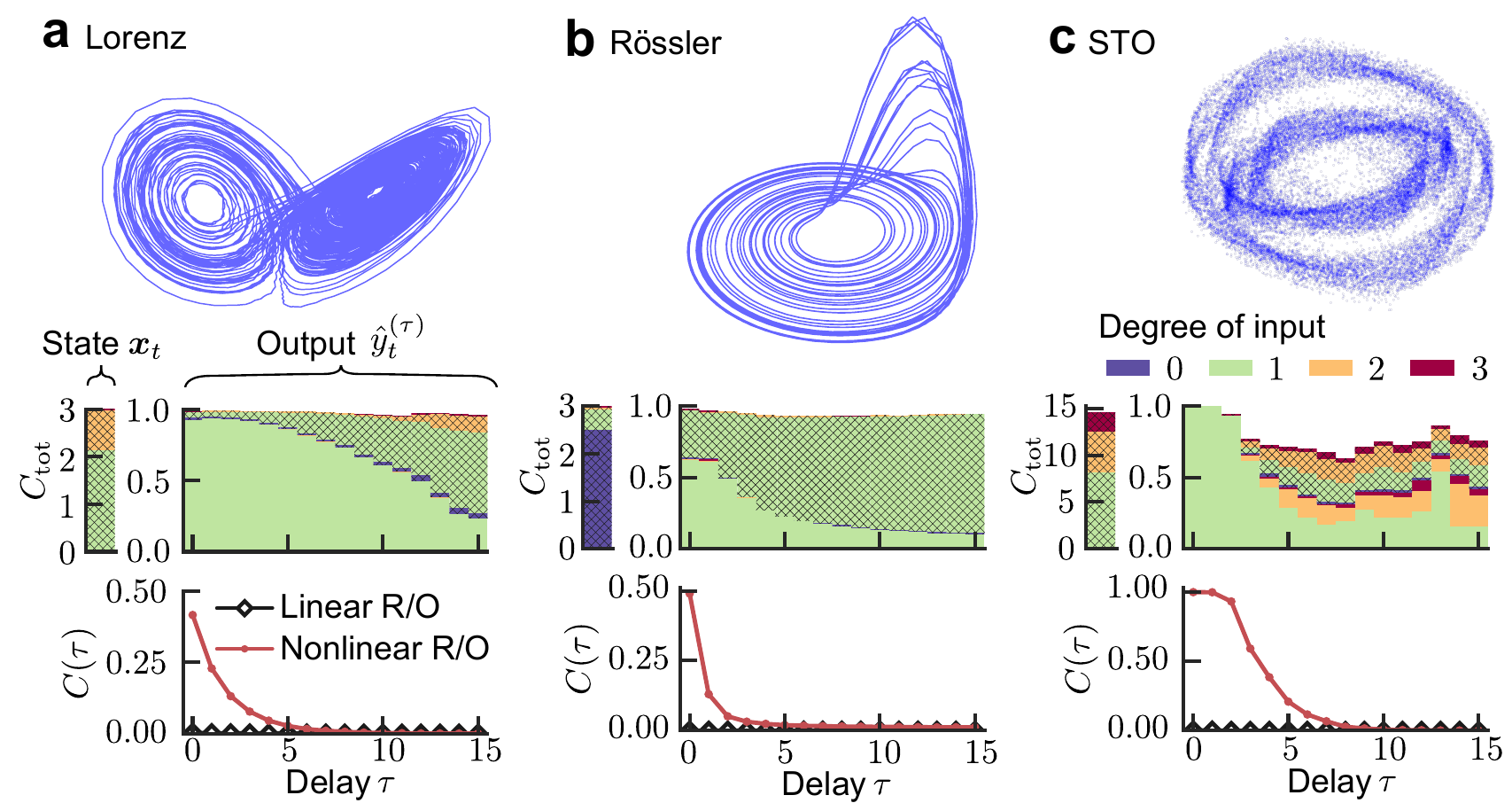}
    \caption{
        \textbf{Memory in systems without ESP}. \textbf{a}, Lorenz model, \textbf{b}, R\"{o}ssler model, and \textbf{c}, real STO. 
        \textbf{a}--\textbf{c} illustrate a 3D-trajectory driven by input (upper), the TIPC decomposition of state and outputs (middle), and memory functions of a dynamical system without ESP (lower). 
        \textbf{c} upper, The time-multiplexing technique was applied to the STO reservoir to make $100$ virtual nodes, whose first three nodes were plotted. 
        The TIPC decomposition of state $\bx_t$ and outputs of a system that trained $\tau$-delayed input $u_{t-\tau}$, using the same colors for the degree of input as in Fig.~\ref{fig:TI_transformation}. 
        The memory functions $C(\tau)$ with linear (black) and nonlinear readout (red). 
        The memory capacities ${\rm MC} = \sum_\tau C(\tau)$ with nonlinear readout are $0.97$ (Lorenz), $0.97$ (R\"{o}ssler), and $4.3$ (STO). 
        See Fig.~S2 for the relationship between the memory function and TIPC. 
    }
    \label{fig:memory}
\end{figure*}

To show that our framework can leverage latent memory in simulated and physical systems without ESP, we calculated memory functions~\cite{jaeger2001short} with nonlinear readout. 
We adopted two chaotic systems (Lorenz and R\"{o}ssler models) and a real spin-torque oscillator (STO) (Fig.~\ref{fig:memory}, upper row), which receive a random input $u_t$ and emulate $\tau$-step delayed input $y_t^{(\tau)} = u_{t-\tau}$ by the nonlinear MLP readout (See Materials and Methods for detailed settings). 
As shown in Fig.~\ref{fig:memory}, TI transformations can successfully extract past inputs from TV states in all three cases.

\begin{figure*}[tb]
    \centering
    \includegraphics[scale=0.4]{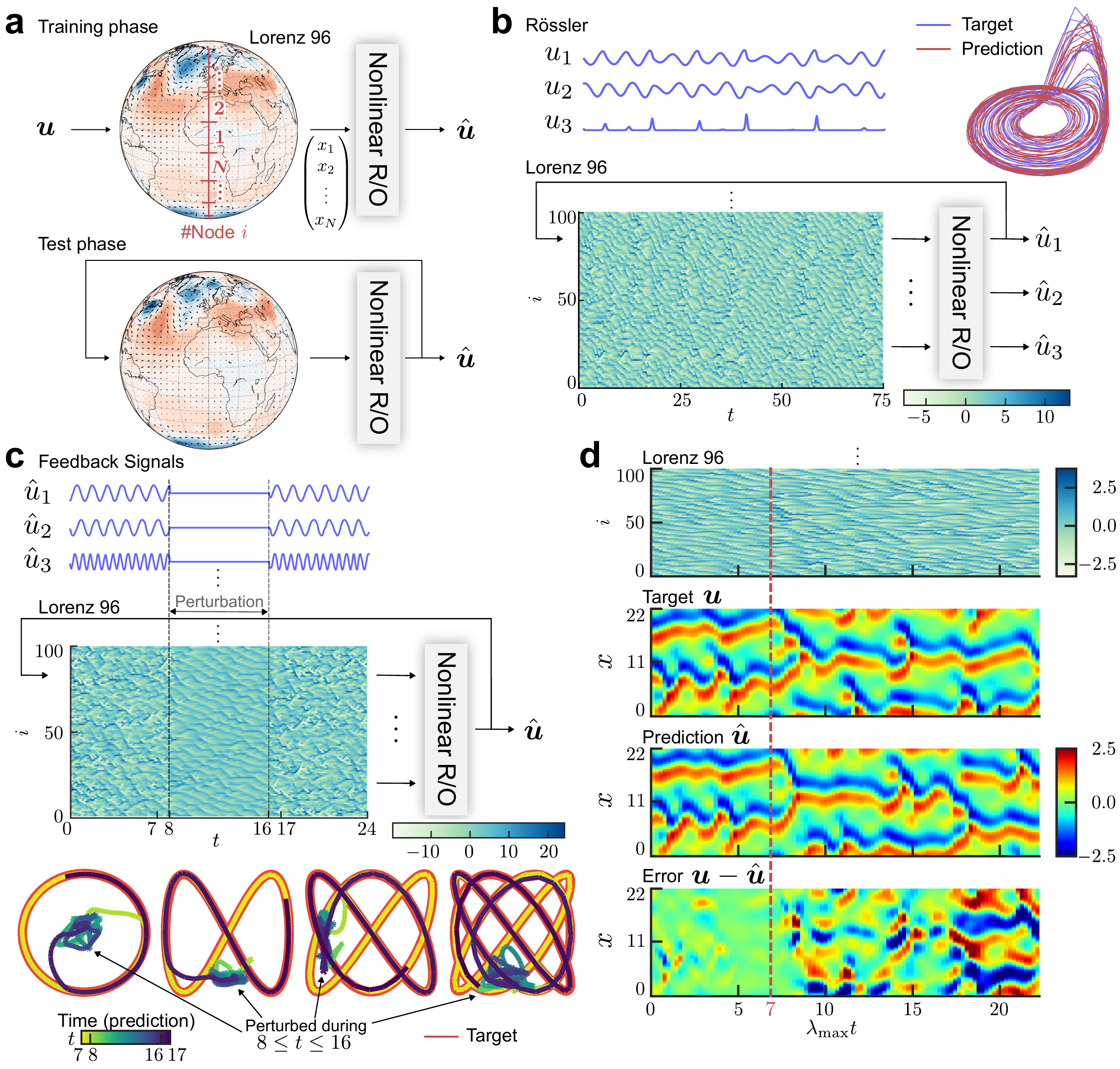}
    \caption{
        \textbf{Application of GRC to the reservoirs without ESP}. 
        \textbf{a}, Procedure of embedding task. 
        Three target systems of 
        \textbf{b}, R\"{o}ssler model, \textbf{c}, Four Lissajous curves, and \textbf{d}, KS model were emulated by the Lorenz 96 model. 
        \textbf{a}, In the training phase, the system receives the current target $\bu$ and predicts the target at the next step by training the nonlinear readout. 
        In the test phase, the input is removed, and the output is fed back into the reservoir as input. 
        \textbf{b}--\textbf{d} illustrate $100$-dimensional time-series of Lorenz 96 out of $500$--$5120$. 
        \textbf{b}, Three variables $(u_1, u_2, u_3)$ of the R\"ossler model (upper left) are the targets, and $u_1$ is the input (lower). 
        The target (blue) and output (red) trajectories are plotted in the 3D space (upper right). 
        After training, the Lorenz 96 exhibited chaos [the maximum conditional Lyapunov exponent in the training phase was $\lambda_{\rm max}=1.18$, and the maximum Lyapunov exponent (MLE) in the test phase was $\lambda_{\rm max}=1.53$]. 
        \textbf{c} The reservoir state (middle) and output $\bu$ (upper) in $0\le t\le24$ are displayed. 
        The feedback outputs were perturbed during $8\le t\le16$. 
        In the test phase, the MLE of the Lorenz 96 reservoir was estimated to be $\lambda_{\rm max}=3.5$.
        Note that the MLE of the embedded signal is estimated to be $\lambda_{\rm max}=0.13$, implying that the embedded attractor is chaotic but has a similar shape as the original one~\cite{kabayama2024designing}. 
        \textbf{d} illustrates the target $\bu$, the prediction $\hat{\bu}$, and their error $\bu-\hat{\bu}$. 
        The MLEs of the target and output were estimated to be $\lambda_{\rm max}=0.13,0.12$, respectively. 
        Note that $\lambda_{\rm max}$ in the horizontal axis represents the maximum Lyapunov exponent of the KS model and is the averaged time length where the initial value's error grows by a factor of $e$. 
    }
    \label{fig:application}
\end{figure*}

\section*{Applications}
Finally, we demonstrate that GRC enables us to construct an information processing system using a dynamical system without ESP. 
Using a high-dimensional atmospheric model of the Lorenz 96 system as a reservoir, which exhibits spatiotemporal chaos, we implemented three types of attractor embedding tasks. 
Figure~\ref{fig:application}a depicts the embedding task in which a reservoir emulates a target dynamical system. 
In the training phase, the reservoir receives the current target state $\bu$ to learn to predict the target at the next step. 
In the test phase, we remove the input and inject its output $\hat{\bu}$ as the next input instead. 
First, to show that the GRC can extract past inputs from a TV state in a temporal task, we embedded the chaotic system of the R\"{o}ssler model, which is described by three variables $\bu=(u_1,u_2,u_3)^\top$. 
In this demonstration, we inject $u_1$ into the Lorenz 96 reservoir for prediction of $\bu$, which requires an input history of $u_1$. 
After training, the Lorenz 96 exhibited chaos; however, it succeeded in reconstructing the rest of the variables $(u_2, u_3)$ (Fig.~\ref{fig:application}b upper right, prediction in red; see Supplementary Information~\SupplSectionAttractorAnalysis and Video \#2 for verification), indicating that the reservoir holds the history of $u_1$. 
Second, to demonstrate that various types of target systems can be emulated by the chaotic system,  we employed four periodic targets simultaneously. 
Figure~\ref{fig:application}c illustrates that four Lissajous curve-shaped trajectories $\hat{\bu}$ with different frequencies were successfully embedded (yellow) in the Lorenz 96 reservoir. 
Even if the feedback signal is fixed at $0$ during $8\le t\le 16$ to perturb the Lorenz 96 state (green to blue), the output returns to the target trajectories (purple). 
Third, to further show the variation of targets in the GRC framework, we employed spatiotemporal chaos, which is the high-dimensional chaos representing complex behavior in nature (e.g., geophysical dynamics and fluid dynamics). 
We used the Kuramoto--Sivashinsky (KS) equation~\cite{kuramoto1976persistent,pathak2018model}, whose state $\bu$ is shown in Fig.~\ref{fig:application}d. 
In the test phase, the Lorenz 96 reservoir predicted $\bu$ and could keep a small error for over seven times the Lyapunov time of the KS model, implying that a similar model is embedded in the reservoir. 
These results suggest that the nonlinear readout allows us to extract memory from complex dynamical systems (e.g., chaos), which are available to solve various types of tasks.

\section*{Discussion}
This paper extended the applicable range of RC to more general systems. 
Conventional RC imposes the ESP condition on the system state to effectively extract computational capabilities. 
The ESP of the state is required to obtain the ESP of output and strongly limits the type of dynamical systems, which leave a wide range of systems behind (e.g., periodic and chaotic dynamics as well as nonstationary ones 
[See Supplementary Information~\SupplSectionTIT for further details]). 
This paper proposed a novel framework by changing the target of the ESP condition from the reservoir state to the final output and by introducing nonlinear readout with memory to utilize the systems without ESP. 
GRC is expected to effectively extract computational capabilities from many types of dynamics without ESP, including physical dynamics, such as spatiotemporal patterns in the brain composed of oscillatory spiking neurons and nonstationary synaptic plasticity.

The components in some types of computers (e.g., transistors in a traditional computer and artificial neurons in a conventional reservoir computer) show identical responses to identical inputs, while real neurons in a biological neural network seem to behave differently in every trial. 
However, living beings should generate meaningful outputs from these signals, which raises a challenge regarding information processing in their nervous systems. 
The proposed framework may provide a novel perspective on this problem. 
Even if the biological system receives an identical input, its neural network does not show identical responses, possibly due to its time dependence, although it certainly processes inputs~\cite{kubota2021unifying}. 
These processed inputs can be transformed through the nonlinear responses of its neural circuit and body, which act as nonlinear readouts with memory, to eventually obtain meaningful outputs.

The use of alternative computing frameworks, such as neuromorphic computing, including PRC or a physical neural network, can take advantage of various properties of physics (e.g., durability and energy efficiency~\cite{torrejon2017neuromorphic}, computational 
speed~\cite{brunner2013parallel,vandoorne2014experimental,larger2017high}, and robustness in extreme experiments, such as radio active ones~\cite{akashi2022coupled}). 
GRC can also exert its power fully when implemented in a physical realm. 
For this purpose, we must develop a nonlinear readout using physical systems with a more refined architecture. 
For example, this can be composed using a physical deep neural network, in which a physical system is utilized as a neural network with training methods~\cite{wright2022deep,nakajima2022physical}. 
In this paper, for the sake of functional verification of GRC, we adopted MLP with backpropagation. 
However, the readout selection is open to various types of functions. 
One approach is to specialize the readout dependent on the type of dynamics. 
For example, detrending for nonstationary systems and envelope extraction for periodic dynamics 
have already been successfully utilized in the post-processing of TV reservoirs, and we can consider these approaches as TI transformations 
(See Supplementary Information~\SupplSectionTIT for further details). 
Another strategy is to employ other nonlinear fitting schemes. 
For example, a dynamical system is also utilized as a readout that can realize TI transformation, holding past inputs, which naturally forms a deep reservoir architecture~\cite{gallicchio2021deep} and can be used for TI transformation. 
If deep physical reservoir architectures are introduced, some reservoirs can be considered to act as readouts of other connected reservoirs. 
This form of network is frequently found in nature. 
Its parameters can be tuned by a gradient-free learning scheme called augmented direct feedback alignment~\cite{nakajima2022physical} if the precise model of physical reservoir is not obtained. 
Further studies on readout functions will open up more sophisticated computational systems.

The proposal of GRC paves the way to exploiting TV terms that have been missing to date. 
In the conventional RC, the reservoir state is required to be TI and can be utilized for computation with a linear or nonlinear readout. 
The computational capability retrievable with linear readout is limited~\cite{dambre2012information,kubota2021unifying}, whereas nonlinear readout eases this restriction. 
First, nonlinear readout can extract infinite past input. 
For instance, binary conversion can theoretically extract all the past inputs from a linear system with binary input~\cite{jaeger2001short}. 
Second, nonlinear readout can make any degree of nonlinearity on the input. 
If the readout has a universal approximation property~\cite{hornik1989multilayer}, the past inputs can be converted into any function. 
In addition to the conventional properties, the mechanism of GRC revealed that the nonlinear readout allows us to utilize not only TI but also TV terms, which contribute strongly to a great variety of feasible outputs (Figs. \ref{fig:TI_transformation} and \ref{fig:memory}).

\section*{Materials and Methods}
\subsection*{Memory Capacity Task}
To evaluate the amount of past inputs held in dynamical systems, we performed the memory capacity task~\cite{jaeger2001short} using both linear and nonlinear readouts. 
We applied a uniform random input $u_t\in[-1,1]$ and a binary random input $u_t\in\{-1,1\}$ to the two chaotic models (Lorenz and R\"ossler models) and the STO, respectively, and set the target output to $\tau$-delayed input $u_{t-\tau}$. 
Using the dynamical state $\bx_t$, we trained the readout weight to form output $\hat{y}_t$ such that the normalized mean-square error (NMSE) $\sum_{t=1}^{T_{\rm train}} (y_t-\hat{y}_t)^2/\sum_{t=1}^{T_{\rm train}} y_t^2$ was minimized. 
The memory function was evaluated by the emulation error $C(\tau)=1-\sum_{t=T_{\rm train}+1}^{T_{\rm train}+T_{\rm test}} (y_t-\hat{y}_t)^2/\sum_{t=T_{\rm train}}^{T_{\rm train}+T_{\rm test}} y_t^2$, and the memory capacity is ${\rm MC}=\sum_\tau C(\tau)$.

\subsection*{Multi-Layer Perceptron}
We employed multi-layer perceptron (MLP) to search for the TI transformation of the chaotic and physical systems. 
MLP uses the nonlinear activation function of the rectified linear unit (ReLU). 
We normalized the input $\bx_t$ to $\hat{\bx}_t$, with zero mean and unit variance, and subsequently injected $\hat{\bx}_t$ to the MLP. 
In the NARMA10 task and memory capacity task for the STO, the $N_i$-dimensional nodes $\hat{\by}_t^{(i)}$ in the $i\uth$ layer are described by 
\begin{align*}
    \hat{\by}^{(i)}_t = \begin{cases}
        \hat{\bx}_t & (i=0) \\
        \bR(\bW^{(i)} \hat{\by}^{(i-1)}_t + \bb^{(i)}) & (i=1,\ldots,L-1) \\
        \bW^{(L)} \hat{\by}^{(L-1)}_t + \bb^{(L)} & (i=L)
    \end{cases}, 
\end{align*}
where $\bW^{(i)}\in\mathbb{R}^{N_i\times N_{i-1}}$ is the weight matrix and $\bb^{(i)}\in\mathbb{R}^{N_i}$ is the bias vector; 
$\bR(\bz)=(R(z_1)\cdots R(z_N))^\top$ and $R(\cdot)$ is the ReLU activation function: 
\begin{align*}
    R(z_j) = \begin{cases}
        z_j & (z_j>0) \\
        0 & (z_j\le 0) 
    \end{cases}. 
\end{align*}
We used four-layer MLPs ($L=4$), and the numbers of nodes were $N_i\in[128,512]~(i=1,2,3)$. 
In the memory capacity tasks of the Lorenz and R\"ossler models and attractor embedding tasks, we added skip connections for each layer, which worked as linear activation functions. 
The nodes are described by 
\begin{align*}
    \hat{\by}_t^{(i)} &= \begin{cases}
    \hat{\bx}_t & (i=0) \\
    \bR(\bW^{(i)} \hat{\by}_t^{(i-1)} + \bb^{(i)}) + 
    \bW_{\rm skip}^{(i)}\hat{\by}_t^{(i-1)} + \bb_{\rm skip}^{(i)} & (i=1, \ldots,L-1) \\
    \bW^{(L)} [\hat{\by}_t^{(L-1)\top} \cdots \hat{\by}_t^{(0)\top}]^\top + \bb^{(L)} & (i=L)
    \end{cases}, 
\end{align*}
where $\bW_{\rm skip}^{(i)}\in\mathbb{R}^{N_i\times N_{i-1}}$ is the weight matrix and $\bb_{\rm skip}^{(i)}\in\mathbb{R}^{N_i}$ is the bias vector. 
We used four- or five-layer MLPs ($L=4,5$), and the numbers of nodes were $N_i\in[64,5120]$ ($i=1,\ldots,L-1$). 
In the training, the batch size was $128$, the maximum epoch was 1,000, and weights with the best performance were selected. 
We used Adam optimizer with a learning rate of $0.001$ and the loss function of the NMSE.

\subsection*{Memories in Systems without ESP}
We calculated the memory functions of the Lorenz and R\"{o}ssler models and the real STO (Fig.~\ref{fig:memory}, upper row) with the nonlinear readout. 
We applied the random input $u_t$ to the systems and emulated $\tau$-step delayed input $y_t^{(\tau)}=u_{t-\tau}$ by the nonlinear MLP readout. 
The lower row in Fig.~\ref{fig:memory} illustrates the memory functions with nonlinear readout (red) as well as those with linear readout (black). 
The conventional linear readout cannot extract memory from these systems at all (the memory capacity ${\rm MC}=\sum_{\tau}C(\tau)=0$ for each system), while the nonlinear MLP readouts successfully recovered latent memories [${\rm MC}=0.97$ (Lorenz), ${\rm MC}=0.97$ (R\"ossler), and ${\rm MC}=4.3$ (STO)]. 
The TIPC elaborates the processed inputs in state and output, revealing the amount and type of processed inputs (middle row in Fig.~\ref{fig:memory}). 
In all three systems, the TIPC decomposition of state was composed of only TV terms. 
Note that the TIPCs were truncated with thresholds to remove numerical errors due to the finite length of the time series, implying that capacities smaller than the threshold potentially exist. 
Conversely, the decomposition of output includes TI terms. 
For each $\tau$, the amount of first-order TI capacity (non-hatched green) was larger than the memory function with nonlinear readout, indicating that the output included not only $u_{t-\tau}$ but also inputs with another delay $u_{t-s}~(s\neq \tau)$ (See Fig.~S2 for further details). 
These results imply that the MLP readouts numerically succeeded in removing time dependence from TV memories and/or in amplifying very small TI memories.

\subsection*{Prerequisite for RC}
In this article, we employed the ESP to represent the condition of conventional RC. 
RC was proposed by integrating two types of recurrent neural networks: an ESN~\cite{jaeger2001echo} and a liquid state machine (LSM)~\cite{maass2002real}, which were independently developed but have similar prerequisites. 
The ESN imposes the network state on the ESP, in which the state is a function of only input history, as follows: 
\begin{align*}
    \bx_t = \bh(u_{t-1},u_{t-2},\ldots). 
\end{align*}
The LSM requires that the state is approximated by a polynomial expansion of input history called the Volterra series. 
The state is expanded by polynomials of input history whose maximum degree is infinite, as follows: 
\begin{align*}
    \bx_t = \ba + \bb_1 u_{t-1} + \bb_2 u_{t-2} + \cdots + \bc_{1} u_{t-1}^2 + \bc_{2} u_{t-1}u_{t-2} + \cdots, 
\end{align*}
where $\ba,\bb_i,\bc_i\in\mathbb{R}^N~(i=1,2,\ldots)$ are the coefficient vectors of zeroth-, first-, and second-order polynomials, respectively. 
Both prerequisites require that the state is independent of time and dependent on input history. 
To straightforwardly define the time-invariance of the state, the ESP was selected in this article.

A related concept to these prerequisites is the generalized synchronization~\cite{lu2018attractor}, in which the dynamical state is synchronized with input. 
Under the circumstance, the state is a function of only input and thus is independent of time. 
The maximum conditional Lyapunov exponent (MCLE) is a measure to judge whether the state is synchronized with input or not. 
A negative MCLE means that the state is a function of only input and thus holds the ESP. 
Conversely, a positive MCLE implies that the state is not a function of only input and does not hold the ESP.

\subsection*{NARMA10 Task}
To demonstrate the TI transformation with the nonlinear data fitting technique, we solved the NARMA10 benchmark task using the ESN~\cite{jaeger2001echo} with TV states, 
where the spectral radius is controlled (set at $1.3$) so that the ESP is broken down. 
The MCLE is estimated to be $\lambda_{\rm max}=1.5\times10^{-2}$ (See Supplementary Information \SupplSectionESN for further details), 
which implies that the states did not hold the ESP~\cite{lu2018attractor}. 
The state $\bx_t=(x_{1,t},\ldots,x_{N,t})^\top$ of the ESN is updated by 
\begin{align}
    x_{i,t+1} = \tanh\left( \sum_{j=1}^N w_{ij}x_{j,t}+w_{{\rm in},i}u_t \right), \label{eq:esn} 
\end{align}
where $u_t$ denotes the uniform random input in the range of $[-1,1]$; 
the input weights $w_{{\rm in},i}$ were generated from a uniform random number in the range of $[-0.1,0.1]$; 
the internal weight matrix ${\bm W}=[w_{ij}]$ was also generated from a uniform random number in $[-1,1]$ and then was rescaled such that its spectral radius is equivalent to $\sigma$. 
In the results of Fig.~\ref{fig:TI_transformation}c, we used $N=100$ and $\sigma=1.3$.

The $10^{\rm th}$-order nonlinear autoregressive moving average (NARMA10) model~\cite{atiya2000new} is described by 
\begin{align*}
    y_{t+1} &= \alpha y_t + \beta y_t \sum_{k=0}^{9} y_{t-k} + \gamma v_t v_{t-9} + \delta, \\
    v_{t} &= \sigma (u_{t} + 1) / 2,
\end{align*}
where $y_t$ is the target output at the $t^{\rm th}$ step; 
$u_t$ is the uniform random input in the range of $[-1, 1]$, which is injected into the reservoir, and $v_t$ linearly converts the range of $u_t$ to $[0, \sigma]$; 
and $(\alpha,\beta,\gamma,\delta,\sigma) = (0.3,0.05,1.5,0.1,0.45)$. 
Note that the target output $y_t$ is a TI function mainly composed of linear past inputs $u_{t-\tau}~(\tau=1,2,3,10,11,12)$ and quadratic past inputs $u_{t-\tau}u_{t-\tau-9}~(\tau=1,2,3)$~\cite{kubota2021unifying}, which represent the input terms required to solve the task.

\subsection*{Attractor Embedding Task}
The embedding tasks were performed with three types of target systems: the R\"ossler model, Lissajous curves, and the KS model. 
In the training phase, we fed the current state of the target system as input $\boldsymbol{u}(t)$ into the Lorenz 96 model and trained a readout weight to emulate the next state of target system $\boldsymbol{u}(t+\Delta t)$.

The Lorenz 96 model, which exhibits spatio-temporally chaotic behavior, was utilized as a reservoir to demonstrate the embedding tasks. 
The differential equation for the $i^{\rm th}$ state $x_i(t)$ is described by 
\begin{align*}
    \frac{dx_i}{dt} = (x_{i+1}-x_{i-2})x_{i-1}-x_i+\mu+\iota u_i~(i=1,\ldots,N),
\end{align*}
where the states are cyclically ordered (i.e., $x_{-1}=x_{N-1}$, $x_0=x_N$, and $x_{N+1}=x_{1}$); 
$\mu$ and $\iota$ are the bias and input intensity, respectively, and $(\mu,\iota)$ were set to $(5, 5)$, $(8, 20)$, and $(5, 15)$ for the embedding tasks of the R\"{o}ssler model, Lissajous curves, and the KS model, respectively; 
in the training phase of embedding task, $u_i(t)$ was the state of the target system, while in the test phase, $u_i(t)$ was the output of MLP fed back as input. 
In the training phase, the MCLEs of the Lorenz 96 model with the target of R\"{o}ssler model, Lissajous curves, and the KS model were estimated to be $\lambda_{\rm max}=1.18,3.67$, and $2.00$, respectively (See Supplementary Information \SupplSectionAttractorAnalysis for further details). 
The positive MCLEs imply that the reservoir states do not hold the ESP~\cite{lu2018attractor}.

We used different settings of input and target for each task. 
In the first task, the target R\"ossler model has the three-dimensional state $(X,Y,Z)$. 
We used only the one-dimensional state $u=X$ for input and target. 
In the second task with the Lissajous curves and the third task with the KS model, all of the five- and $64$-dimensional states, respectively, were used as input and target.

\subsection*{Models}
\subsubsection*{Periodic oscillator}
The equation of the periodic oscillator is described on the polar coordinate. 
The radius $r_t$ and angle $\theta_t$ are updated at each time step as follows: 
\begin{align}
    \begin{split}
        r_{t+1} &= \rho r_t + \mu + \sigma u_{t}, \\
        \theta_{t+1} &= \theta_t + \omega, 
    \end{split} \label{eq:periodic_oscillator}
\end{align}
where $\rho$ is the time constant of relaxation, $\mu$ is the input bias, $\sigma$ is the input intensity, and $\omega$ is the constant angular velocity of $\theta_t$. 
Assuming that $r_0 = \theta_0 = 0$ and $t\rightarrow\infty$, the analytical solution of Eq. (\ref{eq:periodic_oscillator}) is given by 
\begin{align*}
    r_t &= \frac{\mu}{1-\rho} + \sigma\sum_{\tau=1}^\infty \rho^{\tau-1} u_{t-\tau}, \\ 
    \theta_t &= \omega t. 
\end{align*}
Therefore, we analytically obtained the TI radius $r_t=r(u_{t-1}, u_{t-2},\ldots)$ and the angle of function of time $\theta_t=\omega t$. 
The position $(X_t, Y_t) = (r_t \cos\theta_t, r_t \sin\theta_t)$ on the Cartesian coordinate is TV.

\subsubsection*{Lissajous knot}
We demonstrated the second mechanism of the TI transformation using the Lissajous knot~\cite{bogle1994lissajous}, which is a periodic system with multiple frequencies, as follows: 
\begin{align*}
    \begin{pmatrix}
        X_t \\
        Y_t \\
        Z_t
    \end{pmatrix} = 
    \begin{pmatrix}
        \cos(\omega t) \\
        \sin(\omega t) \\
        \sin(2\omega t) + \epsilon u_t
    \end{pmatrix}, 
\end{align*}
where $u_t$ was the uniform random input, and the input intensity was set to $\epsilon=10^{-3}$.

\subsubsection*{Lorenz model}
We evaluated the memory capacity of the Lorenz model. 
The three states $(X, Y, Z)$ are described by 
\begin{align*}
    \frac{dX}{dt} &= p(Y-X) + \iota u_t, \\
    \frac{dY}{dt} &= -XZ + rX - Y, \\
    \frac{dZ}{dt} &= XY - bZ, 
\end{align*}
where $(p, b, r) = (10, 8/3, 28)$ are constant parameters, $u_t\in[-1, 1]$ is the uniform random input, and the input intensity was set to $\iota=30$. 
We solved the equations using the fourth-order Runge-Kutta method with a step width of $\Delta t=0.02$.

\subsubsection*{R\"{o}ssler model}
The R\"{o}ssler system was used for memory capacity and the target of embedding tasks. 
The state variables $(X, Y, Z)$ are updated by 
\begin{align*}
    \frac{dX}{dt} &= -Y - Z, \\
    \frac{dY}{dt} &= X + a Y, \\
    \frac{dZ}{dt} &= b + X Z - c Z + \iota u_t,
\end{align*}
where the constant parameters were set to $(a, b, c) = (0.2, 0.2, 5.7)$; $u_t\in[-1, 1]$ is the uniform random input. 
The input intensity was set to $\iota=0.2~(0)$ for the memory capacity (embedding) task, and the equations were solved by the fourth-order Runge--Kutta method with a step width of $\Delta t=0.1~(0.2)$.

\subsubsection*{Lissajous curves}
Four two-dimensional Lissajous curves are given by $(u_i(t), u_{i+1}(t)) ~ (i=1,2,3,4)$, where $u_i(t)$ is described by 
\begin{align*}
    \begin{pmatrix}
        u_1 \\
        u_2 \\
        u_3 \\
        u_4 \\
        u_5
    \end{pmatrix} = 
    \begin{pmatrix}
         \cos( \omega t) \\
         \sin( \omega t) \\
        -\sin(2\omega t) \\
         \sin(3\omega t) \\
        -\sin(4\omega t)
    \end{pmatrix}. 
\end{align*}

\subsubsection*{Kuramoto--Sivashinsky model}
The KS system is defined by the following partial differential equation for the state $u(x,t)$: 
\begin{align}
    \frac{\partial u}{\partial t} + \frac{\partial^2 u}{\partial x^2} + \frac{\partial^4 u}{\partial x^4} + \frac{1}{2} \left( \frac{\partial u}{\partial x} \right)^2 = 0 \label{eq:kuramoto_sivashinsky}
\end{align}
on a periodic domain $0\le x \le L$ [i.e., $u(x,t)=u(x+L,t)$]. 
We evenly spanned the space to define the $Q$ variables 
\begin{align*}
    \bu(t) = \left(u(\Delta x, t), u(2\Delta x, t), \ldots, u(Q\Delta x, t) \right)^\top
\end{align*}
with an interval of $\Delta x=L/Q$ and numerically solve Eq. (\ref{eq:kuramoto_sivashinsky}) with a step width of $\Delta t=0.25$. 
Note that $L = 22$ and $Q = 64$. 
Using the Rosenstein algorithm~\cite{rosenstein1993practical}, we estimated the MLE of the KS model as $\lambda_{\rm max}=0.13$.

\subsection*{Spin Torque Oscillator}
STO is a device that converts nonlinear spin dynamics into electrical signals. 
We performed PRC on a system in which the device was fed back its own delayed signal. 
The device and feedback circuit were almost the same as those in a previous report~\cite{tsunegi2023information}. 
The delay time of the circuit was about 29 ns, and the feedback gain was about 20 dB. 
The uniform random input $u_t$ was injected
through modulation of the driving voltage of the STO. 
The modulation amplitude and offset were 75 mV and 225 mV, respectively. 
The driving voltage was kept at 20 ns at each step of the input. 
Simultaneously, the STO signal was measured by an oscilloscope with a sampling rate of 5 Gsam/s. 
The measured voltage was treated as 100 virtual nodes, using the time-multiplexing method~\cite{appeltant2011information}.

\subsection*{Temporal Information Processing Capacity}
To reveal the TI and TV representations in the states and outputs, we adopted TIPC~\cite{kubota2021unifying,kubota2023temporal}, which comprehensively quantifies the amount of processed input in the system. 
A dynamical system with input updates the state through state equation $\bx_{t+1}=\bg(\bx_t,u_t)$, where 
$\xt\in\mathbb{R}^N$ and $u_t\in\mathbb{R}$ are the state and input, respectively. 
We assume that the orthonormalized state $\hat{\bx}_t\in\mathbb{R}^r~(r\le N)$ is a function of time and input history: 
\begin{align*}
    \hat{\bx}_t = \boldF(t,u_{t-1},u_{t-2},\ldots), 
\end{align*}
where $r$-normalized, linearly independent state $\hat{\bm x}_t$ is extracted from the $N$-dimensional state $\xt$ through singular value decomposition, 
and that the state can be completely expanded by orthonormal TI bases $I_i(u_{t-1},u_{t-2},\ldots)$ and TV ones $V_i(t,u_{t-1},u_{t-2},\ldots)$ as 
\begin{align*}
    \hat{\bx}_t = \sum_i \ba_i I_i(u_{t-1},u_{t-2},\ldots) + \sum_i \bb_i V_i(t,u_{t-1},u_{t-2},\ldots), 
\end{align*}
where $\ba_i,\bb_i\in\mathbb{R}^r$ are coefficient vectors. 
$I_i$ and $V_i$ represent the function forms of processed inputs, whose amounts are calculated by their squared norms $C_i^{\rm TI}=||\ba_i||^2$ and $C_i^{\rm TV}=||\bb_i||^2$, respectively. 
The total capacity $\Ctot = \sum_i C_i^{\rm TI} + C_i^{\rm TV}$ holds $\Ctot \le r$ (See Supplementary Information~\SupplSectionTIPC for further details).

We used two types of orthonormal bases. 
First, we used the Legendre polynomial-chaos and sinusoidal terms~\cite{kubota2021unifying} to derive the TIPC of the analytical solutions. 
Second, we adopted the Volterra-Wiener-Korenberg series~\cite{korenberg1988identifying,kubota2023temporal} as the orthonormal polynomial expansion for the TIPC of the numerical solutions.

To visually show the TI and TV representation, we used the TIPC decomposition, which sums up the TIPC for each degree $n$ of input as follows: 
\begin{align*}
    C_{{\rm tot},n}^{\rm TI} &= \sum_{\{i|d_{i}=n\}} C_i^{\rm TI}, \\
    C_{{\rm tot},n}^{\rm TV} &= \sum_{\{i|d_{i}=n\}} C_i^{\rm TV}, 
\end{align*}
where $d_i$ denotes the degree of input in the $i\uth$ basis. 
We illustrated the TIPC decomposition using bar graphs, in which nonhatched and hatched areas show the amounts of TI and TV terms in $\hat{\bx}_t$, respectively.

\subsection*{Lyapunov Exponent}
We calculated the MCLE for ESN [Eq. (\ref{eq:esn})] based on the QR decomposition and the Jacobian matrix $\bJ_t = (\bI - {\rm diag}(\bx_t\circ \bx_t))\cdot\bW^\top$, where $\circ$ represents the Hadamard product. 
For the $t\uth$ timestep, we calculated the QR decomposition $\bQ_t\bR_t = \bJ_t\bQ_{t-1}$ to obtain the time series of $\bR_t$, where $\bQ_t$ is the orthonormal matrix and $\bR_t$ is the upper triangular matrix. 
Subsequently, we calculated the Lyapunov exponent by $\lambda_i = \sum_{t=1}^T \ln|R_{ii,t}| / T$ ($i=1,\ldots,N$).

We also calculated the MCLE and MLE for the Lorenz 96 model using the Jacobian matrix (See Supplementary Information~\SupplSectionAttractorAnalysis for further details).

\subsection*{Node-Wise ESP Index}
To find out whether each node of a dynamical system depends on an initial value or not, we calculated the node-wise ESP index. 
We ran the $N$-dimensional dynamical system twice using two different initial values. 
Let $x^{(1)}_{i,t}$, $x^{(2)}_{i,t}$ be the two system states at the $t\uth$ time step with node id $i(=1,\ldots,N)$. 
We defined the node-wise ESP index $\bar{d}_i$ by the NMSE between $x_{i,t}^{(1)}$ and $x_{i,t}^{(2)}$ as follows: 
\begin{align*}
    \bar{d}_i &= \frac{1}{\sigma_i^{(1)}} \sqrt{\frac{1}{T} \sum_{t=1}^T \left(x^{(1)}_{i,t}-x^{(2)}_{i,t}\right)^2}, 
\end{align*}
where $\sigma_i^{(1)}$ is the standard deviation of $x_{i,t}^{(1)}$ during $t\in[1, T]$.

\bibliography{main}

\begin{acknowledgments}
This study is supported by JSPS KAKENHI Grant No. JP21KK0182 and JP23K18472, by JST CREST Grant No. JPMJCR2014, and by the Cross-ministerial Strategic Innovation Promotion Program (SIP) on the ``Integrated Health Care System'' Grant Number JPJ012425. 
Y. I. was supported by JSPS KAKENHI Grant No. JP23KJ0331. 
\end{acknowledgments}

\section*{Data Availability}
The experimental data of the STO are available from the corresponding author on reasonable request.

\section*{Code Availability}
The Python codes that were used for the analysis  can be requested from the corresponding author.

\renewcommand{\theequation}{S\arabic{equation}}
\renewcommand{\thefigure}{S\arabic{figure}}
\renewcommand{\thetable}{S\arabic{table}}
\renewcommand{\thesection}{S\arabic{section}}
\setcounter{equation}{0}
\setcounter{figure}{0}

\clearpage
\section*{Supplementary Information for Reservoir Computing Generalized}

This supplementary information provides a detailed description of the analysis of computational capability and attractor embedding tasks and gives examples of time-invariant transformation. 

\section{Temporal Information Processing Capacity} 
Temporal information processing capacity (TIPC)~\cite{kubota2021unifying} is a measure that comprehensively quantifies the amount of processed inputs in a time-variant (TV) state.

\subsection*{Definition}
TIPC was developed by extending the information processing capacity (IPC)~\cite{dambre2012information}, which is a computational measure that comprehensively evaluates processed inputs in a TI state. 
We consider a nonlinear dynamical system with input whose state equation is described by 
\begin{align*}
    \bx_{t+1} = \boldf(\bx_t, u_t). 
\end{align*}
Under the condition that the state holds an ESP~\cite{jaeger2001echo}, the system state is a function of only input history: 
\begin{align*}
    \bx_t = \bg(u_{t-1},u_{t-2},\ldots), 
\end{align*}
which holds delayed and nonlinearly transformed inputs. 
To quantify the amount of the processed inputs, we begin with the simplest case of memory capacity~\cite{jaeger2001short}, which evaluates the amount of delayed inputs held in the state. 
We apply an independent and identically distributed (i.i.d.) random input $u_t$ to the system and set a target output $z_t$ to $\tau$-delayed input $u_{t-\tau}$. 
\begin{align*}
    z_t = u_{t-\tau}. 
\end{align*}
The output $\hat{z}_t$ is calculated by linear regression as follows: 
\begin{align*}
    \hat{z}_t = \hat{\bW} \bx_t, ~
    \hat{\bW} = \arg\min_{\bW} \sum_{t} (z_t - \bW\bx_t)^2. 
\end{align*}
The memory function $C(\tau)$ is defined by the emulation accuracy of the past input as follows: 
\begin{align}
    C(\tau) &= 1 - \frac{\sum_{t}(z_t-\hat{z}_t)^2}{\sum_t z_t^2}, \label{eqS:memory_function}
\end{align}
where the second term on the right-hand side is the normalized mean square error (NMSE) between the target output $z_t$ and output $\hat{z}_t$. 
If the state perfectly holds the past input $u_{t-\tau}$, the capacity is $1$; if the state does not hold the input at all, the capacity is $0$. 
Additionally, the memory capacity (MC) is defined by a sum of the memory function: 
\begin{align*}
    {\rm MC} = \sum_{\tau} C(\tau). 
\end{align*}
Note that the input must be i.i.d. random because if the targets correlate with each other, $C(\tau)$ cannot represent the amount of past input held in the state. 
For example, if the state is represented by $x_t=u_{t-1}$ (i.e., the state holds only $u_{t-1}$), the memory function should be $C(\tau)=1~(\tau=1)$ and $C(\tau)=0~(\tau>1)$. 
However, if we apply non-i.i.d. input, such as sine wave input, we obtain a memory function $C(\tau)>0~(\tau>1)$, which does not represent the input held in the state. 
Under the condition of i.i.d. input, the MC has an upper bound of rank $r$, i.e., ${\rm MC}\le r$, where $r$ is the rank of the correlation matrix $\bX^\top\bX$.

The state of a nonlinear dynamical system can hold not only delayed inputs but also nonlinear ones, such as $u_{t-1}^2$ and $u_{t-1}u_{t-2}$. 
In the same manner as the memory capacity, we set the target to a nonlinear function of past inputs. 
Due to the same reason as the memory capacity, these targets must not be correlated with each other, i.e., 
\begin{align}
    \sum_{t} z_{i,t}z_{j,t} = 0 ~ (i\neq j). \label{eqS:inner_product}
\end{align}
To meet this condition, we can use an orthogonal polynomial as a target. 
For example, in the case of a uniform random number $u_t\in[-1, 1]$, the target polynomial is orthogonalized with the Legendre polynomial 
$P_n(u)$ [e.g., $u_{t}$, $u_{t}^2$, and $u_{t}^3$ are transformed into $P_1(u_t)=u_t$, $P_2(u_t)=(3u_{t}^2-1)/2$, and $P_3(u_t)=(5u_{t}^3 - u_{t})/2$, respectively]. 
These targets are univariate polynomials, while a multivariate polynomial (e.g., $u_{t-1}u_{t-2}, u_{t-1}^2u_{t-2},u_{t-1}u_{t-2}u_{t-3},\ldots$) is orthogonalized by a product of univariate orthogonal polynomials $\prod_k P_{n_k}(u_{t-\tau_k})$. 
Using the orthogonal basis $z_{i,t}$, the IPC is defined by 
\begin{align}
    C_i = 1 - \frac{\sum_{t}(z_{i,t}-\hat{z}_{i,t})^2}{\sum_t z_{i,t}^2}, \label{eqS:capacity}
\end{align}
where the output is calculated by linear regression in the same manner as memory capacity. 
The total capacity is defined by 
\begin{align}
    \Ctot = \sum_i C_i, \label{eqS:total_capacity}
\end{align}
which is also bounded by the rank $r$, i.e., $\Ctot\le r$. 
If the state is a function of only input history, the total capacity matches the rank. This property is called the completeness property. 
The source codes of IPC are available online~\cite{kubota2022github}, enabling users to easily evaluate the overall computational capabilities of the TI state.

The IPC connects to a Taylor-like polynomial expansion called the polynomial chaos expansion, in which the state is expanded by multivariate orthogonal polynomials of input. 
For example, in the case of the uniform random input, the orthogonal polynomial is the product of the Legendre polynomial, which is called Legendre polynomial chaos. 
The state is expanded by these bases, which are the same as the targets of IPC $z_{i,t}$. 
The state is expanded by the bases as follows: 
\begin{align}
    \hat{\bx}_t = \sum_i \bc_i \hat{z}_{i,t}, \label{eqS:polynomial_chaos_expansion}
\end{align}
where the time series of $\bx_t\in\mathbb{R}^N$ is orthonormalized to that of $\hat{\bx}_t\in\mathbb{R}^r$ using the singular value decomposition. 
The state matrix $\bX = \left[\bx_1 \cdots \bx_T\right]^\top\in\mathbb{R}^{T\times N}$ is decomposed into $\bX = \bU\bSigma\bV^\top$, and $\bU=\left[\hat{\bx}_1\cdots\hat{\bx}_T\right]^\top\in\mathbb{R}^{T\times r}$ includes the orthonormal state vector $\hat{\bx}_t$. 
Let the time series of $z_{i,t}$ be $\bz_i=[z_{i,1}\cdots z_{i,T}]^\top$, and
the time series of $z_{i,t}$ is normalized to that of $\hat{z}_{i,t}$ such that $\hat{z}_{i,t}=z_{i,t}/||\bz_i||$. 
In Eq.~(\ref{eqS:polynomial_chaos_expansion}), the squared norm of the coefficient vector is equivalent to the IPC, as follows: 
\begin{align*}
    C_i=||\bc_i||^2, 
\end{align*}
which means that the magnitude of the coefficient in the expanded state represents the amount of processed input. 
Note that the target output in the temporal task should be a function of input history, i.e., 

\begin{align*}
    y_t = h(u_{t-1},u_{t-2},\ldots). 
\end{align*}
We can apply the IPC to the target and completely expand its polynomials of past input. 
The IPCs of state refer to the amount of processed input in the state, whereas IPCs of target refer to the required input terms to solve the task.

Even if the state is not TI, the inputs can be processed and held in the state. 
If the state is TV, i.e., 
\begin{align}
    \bx_t = \bg(t, u_{t-1}, u_{t-2}, \ldots), \label{eqS:time-variant_state}
\end{align}
processed inputs are represented not only by input history $\{u_{t-1},u_{t-2},\ldots\}$ but also by time $t$. 
To comprehensively find computational capabilities in the state, we must use not only TI targets but also TV ones. 
The TIPC is defined by the same equation as the IPC [Eq.~(\ref{eqS:capacity})]. 
If the targets span a complete orthogonal system for a TV system, and the state is completely expanded by the target bases, the total capacity reaches the rank in the same manner as the IPC.

In this article, we use two types of bases for the TIPC. 
The first is a combination of the Fourier series bases and the Legendre polynomial chaos. 
The bases are composed of three parts: 
the Legendre polynomial chaos $z_t=\prod_k P_{n_k}(u_{t-\tau_k})$, which are the orthogonal bases of only input history if the input is a uniform random number $u_t\in[-1, 1]$; 
the Fourier series bases $z_t = \cos(\omega_n t)$ and $z_t = \sin(\omega_n t)$ $(n=1,2,\ldots)$, which are time-dependent orthogonal bases; 
and the products of the Legendre polynomial chaos and Fourier series bases $z_t=\prod_k P_{n_k}(u_{t-\tau_k})\cos(\omega_n t)$ and $z_t=\prod_k P_{n_k}(u_{t-\tau_k})\sin(\omega_n t)$ ($n=1,2,\ldots$), which are bases of both time and input history. 
All the bases satisfy the orthogonality as in Eq.~(\ref{eqS:inner_product})~\cite{kubota2021unifying} and work as targets of the TIPC. 
We use this type of basis to calculate the TIPC of analytical solutions, such as in the oscillatory dynamics and the Lissajous knot system.

The second basis is the Volterra-Wiener-Korenberg (VWK) series~\cite{korenberg1988identifying,kubota2023temporal}, which is an orthogonal expansion with polynomials of input and state histories. 
Assuming that the state is TV, as in Eq.~(\ref{eqS:time-variant_state}), the delayed state $\bx_{t-\tau}$ can also  be regarded as a term that includes time-dependent elements. 
The VWK series expands the state by polynomials of past input and state and then orthogonalizes them using the Gram--Schmidt. 
\begin{align*}
    \bx_t &= \ba_{1}^{(1)} u_{t-1} + \ba_{2}^{(1)} u_{t-2} + \cdots + \ba_1^{(2)} u_{t-1}^2 + \ba_2^{(2)} u_{t-1}u_{t-2} + \cdots \nonumber \\
    &+ \bb_1^{(1,1)} u_{t-1}x_{1,t-1} + \bb_2^{(1,1)} u_{t-1}x_{2,t-1} + \cdots + \bb_1^{(2,1)} u_{t-1}^2 x_{1,t-1} + \bb_2^{(2,1)} u_{t-1}u_{t-2} x_{1,t-1} + \cdots \nonumber \\
    &+ \bc_1^{(1)} x_{1,t-1} + \bc_2^{(1)} x_{1,t-2} + \cdots + \bc_1^{(2)} x_{1,t-1}^2 + \bc_2^{(2)} x_{1,t-2}^2 + \cdots \\ 
    &= \sum_i \bgamma_i z_{i,t}, 
\end{align*}
where the first to third rows and the fourth row represent expansions before and after orthogonalization, respectively; 
$\ba_i^{(d)}$ is the coefficient vector of the $i\uth$ term of $d\uth$-order input; 
$\bb_i^{(d_1,d_2)}$ is the coefficient vector of the $i\uth$ product term of $d_1$th-order input and $d_2$th-order state; 
$\bc_i^{(d)}$ is the coefficient vector of the $i\uth$ term of $d\uth$-order state; 
$z_{i,t}$ represents the orthogonalized basis; 
and $\bgamma_i$ is the coefficient vector of $z_{i,t}$. 
These types of bases are used to calculate the TIPC of numerical solutions and time series data, such as the Lorenz model, the R\"ossler model, and real spin-torque oscillator (STO) data.

\begin{figure}[tb]
    \centering
    \includegraphics[scale=0.28]{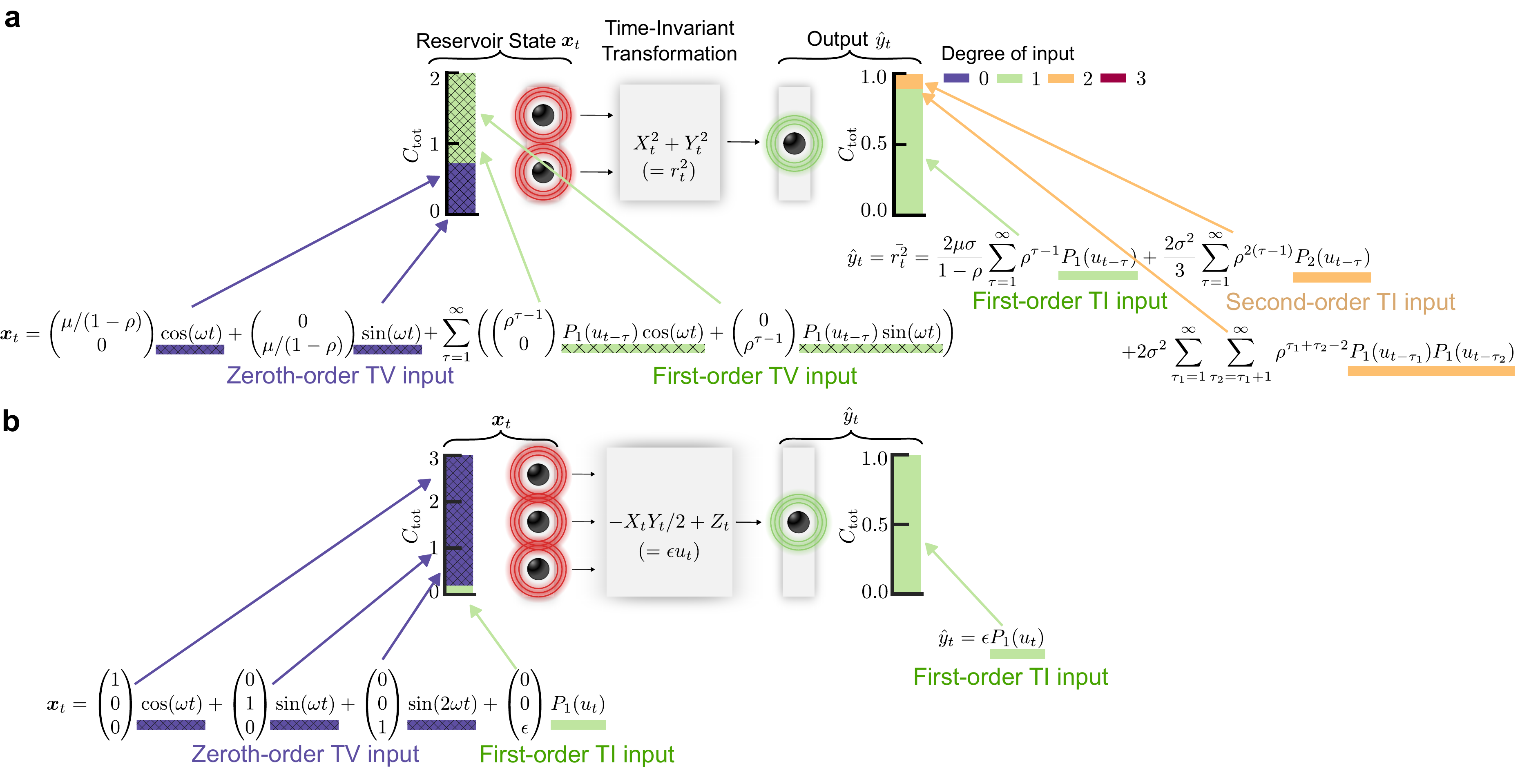}
    \caption{
        \textbf{
        Explanation of TIPC with examples of TI transformation in Fig.~2. 
        } 
        The TI transformations of \textbf{a}, oscillatory dynamics and \textbf{b}, Lissajous knot converts the TV reservoir state $\bx_t$ in Eqs.~(\ref{eqS:oscillatory_dynamics_state_orthogonal_expansion}) and (\ref{eqS:lissajous_knot_state_orthogonal_expansion}) into the TI output $\hat{y}_t$ in Eqs.~(\ref{eqS:oscillatory_dynamics_output_orthogonal_expansion}) and (\ref{eqS:lissajous_knot_output_orthogonal_expansion}), respectively. 
        The state and output are expanded by the TI and TV terms. 
        After normalizing the state and orthogonal bases, the squared norm of coefficient vector represents the TIPC. 
        The TIPCs are summarized as the TIPC decomposition $\Ctot$, which is depicted by color bars, where the non-hatched and hatched bars represent the TI and TV capacities, respectively. 
        The color of the TIPC represents the degree of input: $0$ (purple), $1$ (green), $2$ (orange), and $3$ (red). 
        The orthogonal bases are underlined by the bar corresponding to that of the TIPC decomposition. 
    }
    \label{figS:TIPC}
\end{figure}

\subsection*{Derivation of TIPC of Oscillatory Dynamics}
As shown in Fig.~2a of the main text, we calculate the TIPC of oscillatory dynamics, whose state equation is described by 
\begin{align}
    \begin{split}
        r_{t+1} &= \rho r_t + \mu + \sigma u_{t}, \\
        \theta_{t+1} &= \theta_t + \omega, 
    \end{split} \label{eqS:limit_cycle}
\end{align}
where $r_t$ and $\theta_t$ are the radius and angle on the polar coordinate, respectively, $\rho~(<1)$ is the time constant of relaxation, $\mu$ is the input bias, $\sigma$ is the input intensity, and $\omega$ is the constant angular velocity of $\theta_t$. 
Assuming that $r_0 = \theta_0 = 0$ and $t\rightarrow\infty$, the analytical solution of Eq.~(\ref{eqS:limit_cycle}) is as follows: 
\begin{align*}
    r_t &= \frac{\mu}{1-\rho} + \sigma\sum_{\tau=1}^\infty \rho^{\tau-1} u_{t-\tau}, \\ 
    \theta_t &= \omega t. 
\end{align*}
The reservoir state on the Cartesian coordinate is 
\begin{align*}
    \bx_t = \begin{pmatrix}
        X_t \\
        Y_t
    \end{pmatrix} = 
    \begin{pmatrix}
        r_t \cos\theta_t \\
        r_t \sin\theta_t
    \end{pmatrix} 
    = \left(\frac{\mu}{1-\rho} + \sum_{\tau=1}^\infty \rho^{\tau-1} u_{t-\tau} \right)
    \begin{pmatrix}
        \cos(\omega t) \\
        \sin(\omega t)
    \end{pmatrix}. 
\end{align*}
Using the orthogonal bases, the state can be expanded as follows: 
\begin{align}
    \bx_t = 
    \begin{pmatrix}
        \mu/(1-\rho) \\
        0
    \end{pmatrix} \cos(\omega t)
    + 
    \begin{pmatrix}
        0 \\
        \mu/(1-\rho) 
    \end{pmatrix} \sin(\omega t)
    + \sum_{\tau=1}^\infty \left( 
    \begin{pmatrix} 
        \rho^{\tau-1} \\
        0
    \end{pmatrix} P_1\left(u_{t-\tau}\right) \cos(\omega t) + 
    \begin{pmatrix}
        0 \\
        \rho^{\tau-1}
    \end{pmatrix} P_1\left(u_{t-\tau}\right) \sin(\omega t)
    \right), \label{eqS:oscillatory_dynamics_state_orthogonal_expansion}
\end{align}
where $P_1(u_{t-\tau})=u_{t-\tau}$ is the first-order Legendre polynomial. 
Letting $\bX = (X_1\cdots X_T)^\top$ and $\bY = (Y_1\cdots Y_T)^\top$, we obtain
\begin{align*}
    ||\bX|| &= ||\bY|| = \frac{1}{1-\rho}\sqrt{\frac{T}{6} \left(3\mu^2 + \frac{1-\rho}{1+\rho}\right)}. 
\end{align*}
The normalized state is expanded by the orthonormal bases as follows: 
\begin{align*}
    \hat{\bx}_t &= \begin{pmatrix}
        X_t / ||\bX|| \\
        Y_t / ||\bY|| 
    \end{pmatrix} = \frac{\bx_t}{||\bX||}\\ 
    &= 
    \begin{pmatrix}
        \frac{\mu\sqrt{T/2}}{(1-\rho)||\bX||} \\
        0
    \end{pmatrix} \frac{\cos(\omega t)}{\sqrt{T}\sigma_{{\rm c}}} + 
    \begin{pmatrix}
        0 \\
        \frac{\mu\sqrt{T/2}}{(1-\rho)||\bX||} 
    \end{pmatrix} \frac{\sin(\omega t)}{\sqrt{T}\sigma_{{\rm s}}}
    + \sum_{\tau=1}^\infty
    \begin{pmatrix}
        \frac{\sqrt{T/6}\rho^{\tau-1}}{||\bX||} \\
        0
    \end{pmatrix} \frac{P_1(u_{t-\tau})\cos(\omega t)}{\sqrt{T}\sigma_{u{\rm c}}} \\
    &+ \sum_{\tau=1}^\infty 
    \begin{pmatrix}
        0 \\
        \frac{\sqrt{T/6}\rho^{\tau-1}}{||\bX||}
    \end{pmatrix} \frac{P_1(u_{t-\tau})\sin(\omega t)}{\sqrt{T}\sigma_{u{\rm s}}}. 
\end{align*}
Therefore, the TIPCs for the TV bases $\cos(\omega t)$, $\sin(\omega t)$, $P_1(u_{t-\tau})\cos(\omega t)$, and $P_1(u_{t-\tau})\sin(\omega t)$ are given by 
\begin{align*}
    C_{\cos(\omega t)} &= C_{\sin(\omega t)} = \left(\frac{\sqrt{T/2}\mu}{(1-\rho)||\bX||}\right)^2 = \frac{3\mu^2}{3\mu^2 + (1-\rho)/(1+\rho)}, \\
    C_{P_1(u_{t-\tau})\cos(\omega t)} &= C_{P_1(u_{t-\tau})\sin(\omega t)} = \left(\frac{\sqrt{T/6}\rho^{\tau-1}}{||\bX||}\right)^2 = \frac{\rho^{2(\tau-1)}(1-\rho)^2}{3\mu^2 + (1-\rho)/(1+\rho)}, 
\end{align*}
respectively (Fig.~\ref{figS:TIPC}a). 
Their total capacity is 
\begin{align*}
    \Ctot = C_{\cos(\omega t)} + C_{\sin(\omega t)} + \sum_{\tau=1}^\infty \left(C_{P_1(u_{t-\tau})\cos(\omega t)} + C_{P_1(u_{t-\tau})\sin(\omega t)}\right) = 2, 
\end{align*}
which holds the completeness property. 
In the main text, the example of TI transformation for the periodic oscillator is 
\begin{align*}
    f(\bx_t) = X_t^2 + Y_t^2 = r_t^2. 
\end{align*}
Expanding the output $\hat{y}_t=f(\bx_t)$ by the orthonormal bases of Legendre polynomial chaos, we analytically derive its TIPC. 
The output is expanded by orthogonal linear and quadratic input terms as follows: 
\begin{align*}
    r_t^2 &=  
    \left( \frac{\mu}{1-\rho} + \sigma\sum_{\tau=1}^\infty \rho^{\tau-1} u_{t-\tau}\right)^2 \nonumber \\
    &= \left(\frac{\mu}{1-\rho}\right)^2 + \frac{\sigma^2}{3(1-\rho^2)} 
    + \frac{2\mu\sigma}{1-\rho} \sum_{\tau=1}^\infty \rho^{\tau-1} P_1(u_{t-\tau}) + \frac{2\sigma^2}{3} \sum_{\tau=1}^\infty \rho^{2(\tau-1)} P_2(u_{t-\tau}) \nonumber \\
    &+ 2\sigma^2 \sum_{\tau_1=1}^\infty \sum_{\tau_2=\tau_1+1}^\infty \rho^{\tau_1+\tau_2-2} P_1(u_{t-\tau_1})P_1(u_{t-\tau_2}), 
\end{align*}
where $P_1(u_{t-\tau})=u_{t-\tau}$ and $P_2(u_{t-\tau})=(3u_{t-\tau}^2-1)/2$ are the first- and second-order Legendre polynomials, respectively. 
We calculate the debiased output $\bar{r_t^2}$ by subtracting the constant terms from $r_t^2$: 
\begin{align}
    \bar{r^2_t} &= \frac{2\mu\sigma}{1-\rho} \sum_{\tau=1}^\infty \rho^{\tau-1} P_1(u_{t-\tau}) 
    + \frac{2\sigma^2}{3} \sum_{\tau=1}^\infty \rho^{2(\tau-1)} P_2(u_{t-\tau}) 
    + 2\sigma^2 \sum_{\tau_1=1}^\infty \sum_{\tau_2=\tau_1+1}^\infty \rho^{\tau_1+\tau_2-2} P_1(u_{t-\tau_1})P_1(u_{t-\tau_2}). \label{eqS:oscillatory_dynamics_output_orthogonal_expansion}
\end{align}
Letting $\bar{\br^2} = (\bar{r^2_1}\cdots \bar{r^2_T})^\top$, its norm is 
\begin{align*}
    ||\bar{\br^2}|| = \frac{2\sigma}{3} \sqrt{\frac{T}{1-\rho^2} \left( \frac{3\mu^2}{(1-\rho)^2} + \frac{\sigma^2}{5(1+\rho^2)} +  \frac{\sigma^2\rho}{1-\rho}\right)}. 
\end{align*}
The normalized output is expanded by orthonormal bases as follows: 
\begin{align*}
    \frac{\bar{r^2_t}}{||\bar{\br^2}||} &= \frac{2\mu\sigma}{(1-\rho)||\bar{\br^2}||}\sqrt{\frac{T}{3}} \sum_{\tau=1}^\infty \rho^{\tau-1} \frac{P_1(u_{t-\tau})}{\sqrt{T/3}} 
    + \frac{2\sigma^2}{3||\bar{\br^2}||} \sqrt{\frac{T}{5}} \sum_{\tau=1}^\infty \rho^{2(\tau-1)} \frac{P_2(u_{t-\tau})}{\sqrt{T/5}} \\
    &+ \frac{2\sigma^2\sqrt{T}}{3||\bar{\br^2}||} \sum_{\tau_1=1}^\infty \sum_{\tau_2=\tau_1+1}^\infty \rho^{\tau_1+\tau_2-2} \frac{P_1(u_{t-\tau_1})P_1(u_{t-\tau_2})}{\sqrt{T}/3}. 
\end{align*}
The TIPCs for the bases $P_1(u_{t-\tau})$, $P_2(u_{t-\tau})$, and $P_1(u_{t-\tau_1})P_1(u_{t-\tau_2})~(\tau_1<\tau_2)$ are given by 
\begin{align*}
    C_{P_1(u_{t-\tau})} &= \frac{3\mu^2 (1-\rho^2)}{ 3\mu^2 + \sigma^2(1-\rho)^2/(5(1+\rho^2)) +  \sigma^2\rho(1-\rho)} \rho^{2(\tau-1)}, \\
    C_{P_2(u_{t-\tau})} &= \frac{\sigma^2(1-\rho^2)(1-\rho)^2/5}{3\mu^2 + \sigma^2(1-\rho)^2/(5(1+\rho^2)) +  \sigma^2\rho(1-\rho)} \rho^{4(\tau-1)}, \\
    C_{P_1(u_{t-\tau_1})P_1(u_{t-\tau_2})} &= \frac{\sigma^2(1-\rho^2)(1-\rho)^2}{3\mu^2 + \sigma^2(1-\rho)^2/(5(1+\rho^2)) +  \sigma^2\rho(1-\rho)} \rho^{2(\tau_1+\tau_2-2)}, 
\end{align*}
respectively (Fig.~\ref{figS:TIPC}a).

\subsection*{Derivation of TIPC of Lissajous Knot}
As shown in Fig.~2b of the main text, we calculate the TIPC of the Lissajous knot, whose state is described by 
\begin{align*}
    \bx_t = 
    \begin{pmatrix}
        X_t \\
        Y_t \\
        Z_t
    \end{pmatrix} = 
    \begin{pmatrix}
        \cos(\omega t) \\
        \sin(\omega t) \\
        \sin(2\omega t) + \epsilon u_t
    \end{pmatrix}. 
\end{align*}
The state is expanded by the orthogonal bases as follows: 
\begin{align}
    \bx_t = 
    \begin{pmatrix}
        1 \\
        0 \\
        0
    \end{pmatrix} \cos(\omega t) + 
    \begin{pmatrix}
        0 \\
        1 \\
        0
    \end{pmatrix} \sin(\omega t) + 
    \begin{pmatrix}
        0 \\
        0 \\
        1 
    \end{pmatrix} \sin(2\omega t) + 
    \begin{pmatrix}
        0 \\
        0 \\
        \epsilon
    \end{pmatrix} P_1(u_t), \label{eqS:lissajous_knot_state_orthogonal_expansion}
\end{align}
where $P_1(u_t)=u_t$ is the first-order Legendre polynomial. 
Let $\bX = (X_1\cdots X_T)^\top$, $\bY = (Y_1\cdots Y_T)^\top$, and $\bZ = (Z_1\cdots Z_T)^\top$. 
Their norms are given by 
\begin{align*}
    ||\bX|| = ||\bY|| = \sqrt{T/2}, ~ ||\bZ|| = \sqrt{T(2\epsilon^2+3)/6}.
\end{align*}
The normalized state is expanded by the orthonormal bases as follows: 
\begin{align*}
    \hat{\bx}_t = 
    \begin{pmatrix}
        X_t / ||\bX|| \\
        Y_t / ||\bY|| \\
        Z_t / ||\bZ||
    \end{pmatrix} = 
    \begin{pmatrix}
        1 \\
        0 \\
        0
    \end{pmatrix} \frac{\cos(\omega t)}{\sqrt{T/2}} + 
    \begin{pmatrix}
        0 \\
        1 \\
        0
    \end{pmatrix} \frac{\sin(\omega t)}{\sqrt{T/2}} + 
    \begin{pmatrix}
        0 \\
        0 \\
        \sqrt{3/(2\epsilon^2+3)} 
    \end{pmatrix} \frac{\sin(2\omega t)}{\sqrt{T/2}} + 
    \begin{pmatrix}
        0 \\
        0 \\
        \epsilon \sqrt{2/(2\epsilon^2+3)}
    \end{pmatrix} \frac{P_1(u_t)}{\sqrt{T/3}}
\end{align*}
The TIPCs for the bases $\cos(\omega t)$, $\sin(\omega t)$, $\sin(2\omega t)$, and $P_1(u_{t})$ are given by 
\begin{align*}
    C_{\cos(\omega t)} = C_{\sin(\omega t)} = 1, ~ C_{\sin(2\omega t)} = \frac{3}{2\epsilon^2+3}, ~ C_{P_1(u_{t})} = \frac{2\epsilon^2}{2\epsilon^2+3}, 
\end{align*}
respectively (Fig.~\ref{figS:TIPC}b).

In the main text, the example of TI transformation for the Lissajous knot is 
\begin{align*}
    f(\bx_t) = -2X_t Y_t + Z_t = \epsilon u_t. 
\end{align*}
The output is expanded by an orthogonal term, as follows: 
\begin{align}
    f(\bx_t) &= \epsilon P_1(u_t), \label{eqS:lissajous_knot_output_orthogonal_expansion}
\end{align}
where $P_1(u_{t})=u_{t}$ is the first-order Legendre polynomial. 
The TIPC for the basis $P_1(u_{t})$ is given by 
\begin{align*}
    C_{P_1(u_{t})} &= 1, 
\end{align*}
which is depicted in Fig.~\ref{figS:TIPC}b.

\begin{figure}[tb]
    \centering
    \includegraphics[scale=0.4]{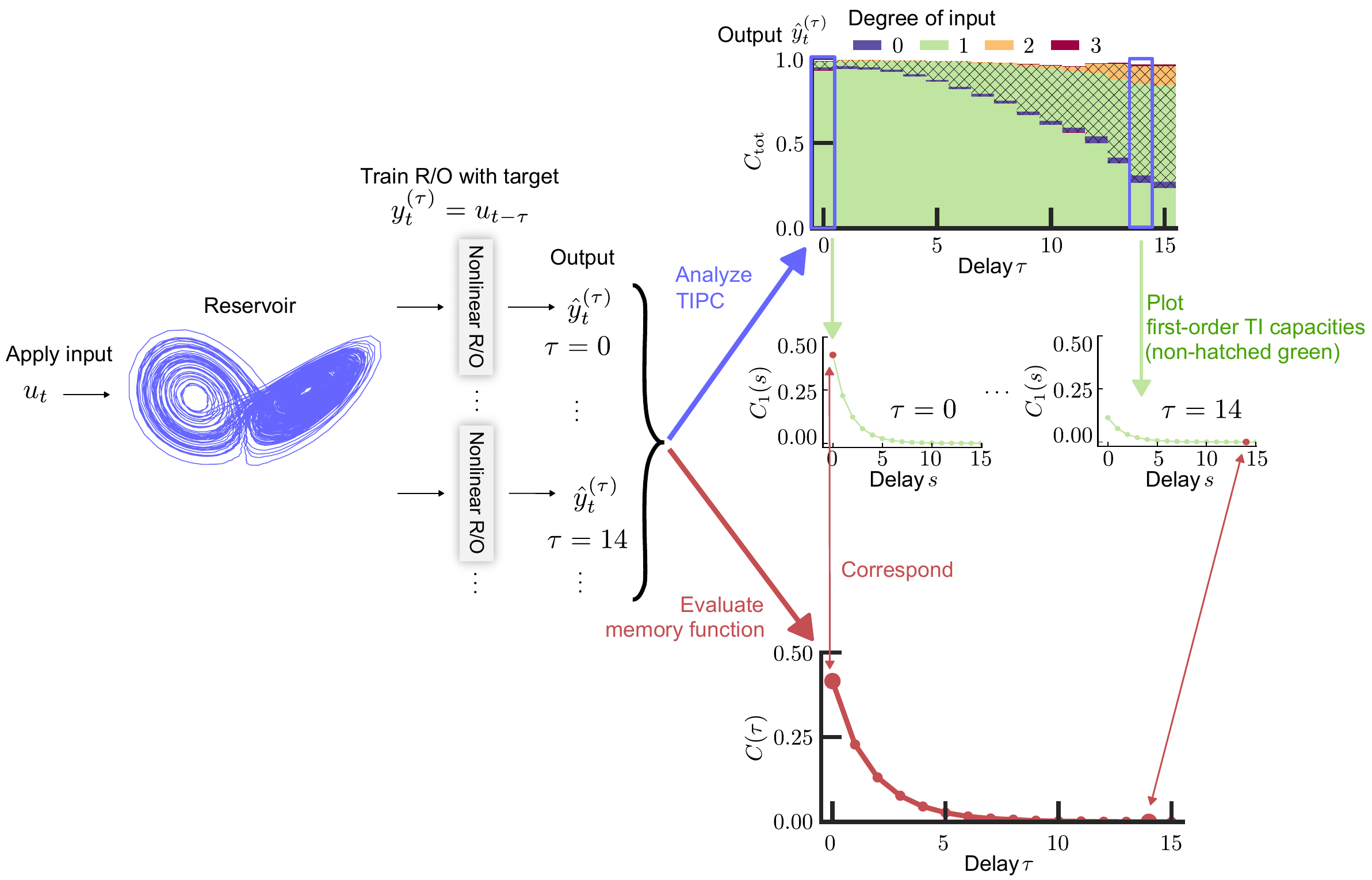}
    \caption{
        \textbf{
        Correspondence between MC and TIPC of output. 
        } 
        A reservoir receives an input $u_t$, and targets $y_t^{(\tau)}=u_{t-\tau}$ ($\tau=0,\ldots,15$) are emulated by nonlinear readout. 
        The memory function is evaluated by $C(\tau)=1-\sum_t(y_t^{(\tau)}-\hat{y}_t^{(\tau)})^2 / \sum_t(y_t^{(\tau)})^2$ (bold red arrow). 
        The outputs are also analyzed by the TIPC (blue arrow). 
        The TIPC decomposition $\Ctot$ is depicted by color bars, where the non-hatched and hatched bars represent the TI and TV capacities, respectively. 
        The color of the TIPC represents the degree of input: $0$ (purple), $1$ (green), $2$ (orange), and $3$ (red). 
        The TIPC includes the TI first-order capacity $C_1(s)$, which is the capacity of $u_{t-s}$. 
        Its sum is represented by the non-hatched green bar in the TIPC decomposition. 
        In the middle-right position, $C_1(s)$ with $\tau=0,14$ is plotted. 
        The TI first-order capacity $C_1(s)$ with $s=\tau$ corresponds with the memory function $C(\tau)$ (thin red arrows). 
    }
    \label{figS:Correspondence}
\end{figure}

\subsection*{Correspondence Between MC and TIPC}
As shown in Fig.~3 in the main text, we demonstrate that, even if the reservoir holds only the TV terms, we can extract the TI terms using the nonlinear readout. 
We inject the input $u_t$ into the TV reservoir and emulate the target $y_t^{(\tau)}=u_{t-\tau}$, whose accuracy is evaluated by the memory function $C(\tau)$, as in Eq.~(\ref{eqS:memory_function}) (bold red arrow). 
We also analyze the TIPC of the output (blue arrow), which represents the magnitude of the coefficient of the polynomial-expanded output. 
As shown in the upper-right position of Fig.~\ref{figS:Correspondence}, the TIPC decomposition $\Ctot$ includes both the TI and TV capacities. 
We extract the TI first-order capacities $C_1(s)$, which represent the amount of $u_{t-s}$ in the output $\hat{y}_t^{(\tau)}$, and plot them as the forgetting curve (green arrow) in the middle-right position. 
We plot $C_1(s)$ with $\tau=0,14$, whose red point is $C_1(s)$ with $s=\tau$ and corresponds with the memory function $C(\tau)$ (thin red arrows). 
Since the non-hatched green bar in the TIPC decomposition is the sum of $C_1(s)$, it includes the memory function $C(\tau)$.

\section{Node-Wise Echo State Property} \label{sec:node_wise_esp}
We consider a dynamical system with $N$-dimensional states $\bx_t=(x_{1,t}\cdots x_{N,t})^\top\in\mathbb{R}^N$ that receives $M$-dimensional input $\bu_t\in\mathbb{R}^M$, as follows: 
\begin{align*}
    \bx_{t+1} = \bg(\bx_t, \bu_{t}), 
\end{align*}
where $\bg: \mathbb{R}^N\times\mathbb{R}^M\rightarrow\mathbb{R}^N$. 
We assume that the state is expanded by the TI function $\bI$ and TV function $\bV$~\cite{kubota2021unifying}, as follows: 
\begin{align*}
    \bx_{t} = \bI(\bu_{t-1}, \bu_{t-2}, \ldots) + \bV(t, \bu_{t-1}, \bu_{t-2}, \ldots), 
\end{align*}
where $\bI=(I_1\cdots I_N)^\top$ and $\bV=(V_1\cdots V_N)^\top$. 
To obtain outputs with an echo state property (ESP), one or more nodes must have the ESP, under which the state is independent of time. 
We say that the $i$th state has a node-wise ESP if the following relation is held: 
\begin{align*}
    x_{i,t} = I_i(\bu_t,\bu_{t-1},\ldots). 
\end{align*}
Under the assumption that one or more nodes have node-wise ESP, linear regression can form time-invariant (TI) outputs: 
\begin{align}
    \hat{\by}_{t} = \sum_{i\in S_I} \bw_i x_{i,t} = \sum_{i\in S_I} \bw_i I_i(\bu_t,\bu_{t-1},\ldots), \label{eqS:output_with_TI_state}
\end{align}
where $S_I$ is an index set whose node is TI, 
and $\bw_i(\neq\boldsymbol{0})$ is the weight vector for the $i$th node. 
If the node does not have node-wise ESP, the $i$th state is described by 
\begin{align*}
    x_{i,t} = I_i(\bu_t,\bu_{t-1},\ldots) + V_i(t, \bu_t,\bu_{t-1},\ldots). 
\end{align*}
If none of the states have node-wise ESP, the outputs are described by 
\begin{align}
    \hat{\by}_t = \sum_{i\in S_V} \bw_i x_{i,t} = \sum_{i\in S_V} \bw_i I_i(\bu_t,\bu_{t-1},\ldots) + \sum_{i\in S_V} \bw_i V_i(t, \bu_t,\bu_{t-1},\ldots). \label{eqS:output_with_TV_state}
\end{align}
where $S_V$ is an index set whose node is TV, 
and $\bw_i(\neq\boldsymbol{0})$ is the weight vector for the $i$th node. 
Even if none of the states have node-wise ESP, the output can be TI in a special case. 
Under the condition that the TV elements $V_i~(i\in S_V)$ are linearly dependent from each other, i.e., 
\begin{align*}
    V_{i} = \sum_{\substack{j\in S_V \\ j\neq i}} a_j V_j, 
\end{align*}
the output can form TI functions, as follows: 
\begin{align*}
    \hat{\by}_t = \bw \left( x_{i,t} - \sum_{\substack{j\in S_V \\ j\neq i}} a_i x_{i,t} \right) = \bw \left( I_i(\bu_t,\bu_{t-1},\ldots) - \sum_{\substack{j\in S_V \\ j\neq i}} a_j I_j(\bu_t,\bu_{t-1},\ldots) \right), 
\end{align*}
where $a_i\in\mathbb{R}$ and $\bw\in\mathbb{R}^N$. 
Under the conditions that none of the states have node-wise ESP and the TV elements $V_i$ are linearly independent from each other, the outputs in Eq.~(\ref{eqS:output_with_TV_state}) cannot form TI functions.

\begin{figure}[tb]
    \centering
    \includegraphics[scale=1.0]{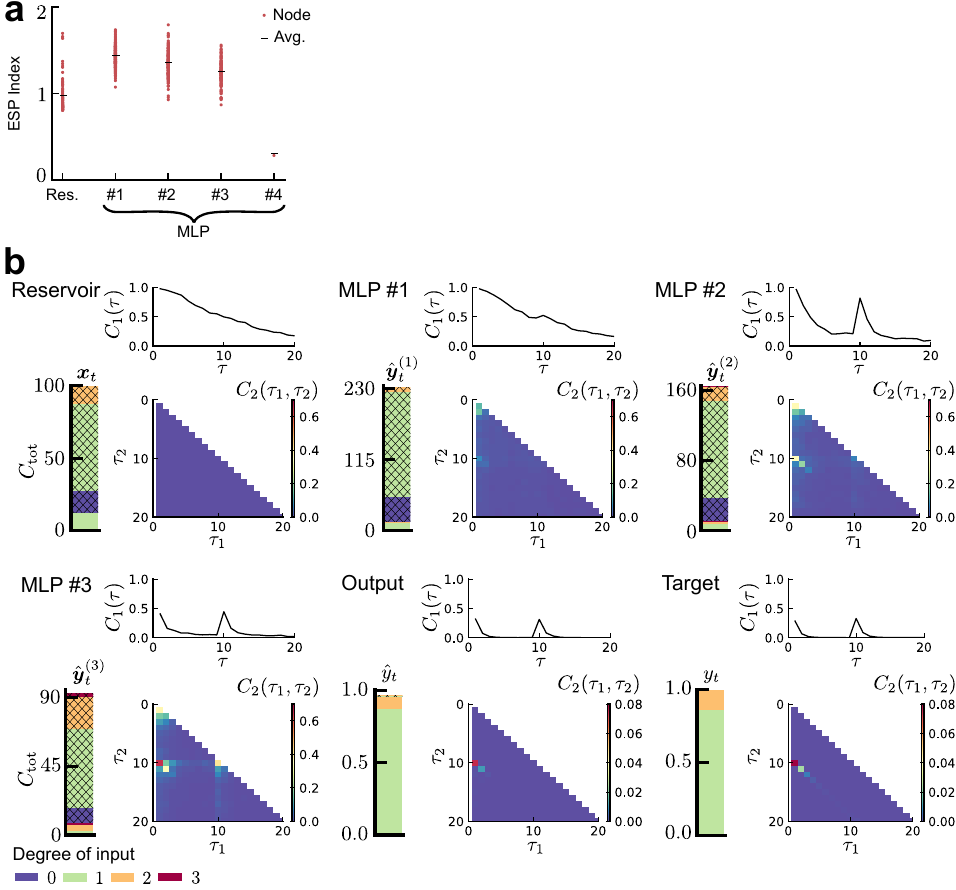}
    \caption{
        \textbf{
        ESP Index and TIPC of ESN with MLP. 
        } 
        \textbf{a}, Node-wise ESP indices of ESN with 4-layer MLP.         
        The vertical axis is the node-wise ESP index, and the horizontal axis is the layer name, in which ``Res.'' represents the reservoir layer. 
        The numbers of nodes are $(100, 256, 192, 128, 1)$ from Res. to \#4. 
        \textbf{b}, The TIPC decomposition (lower left) and its TI first-order capacity $C_1(\tau)$ (upper right; non-hatched green in the TIPC) and TI second-order capacity $C_2(\tau_1,\tau_2)$ (lower right; non-hatched orange in the TIPC) for each layer. 
        $C_1(\tau)$ and $C_2(\tau_1,\tau_2)$ represent the capacities of orthogonalized $u_{t-\tau}$ and $u_{t-\tau_1}u_{t-\tau_2}$ ($\tau_1\le\tau_2$), respectively. 
        The lower rightmost plots illustrate the capacities required to emulate the NARMA10 target. 
        The TIPC decomposition $\Ctot$ is depicted by color bars, where the non-hatched and hatched bars represent the TI and TV capacities, respectively. 
        The color of the TIPC represents the degree of input: $0$ (purple), $1$ (green), $2$ (orange), and $3$ (red). 
    }
    \label{figS:ESN}
\end{figure}

\section{Echo State Network with Multi-Layer Perceptron}
\subsection*{ESP Index and Conditional Lyapunov Exponent}
In Fig.~2c of the main text, to examine whether nodes in each layer have ESP or not, we calculated the node-wise ESP index of the echo state network (ESN) with 4-layer multi-layer perceptron (MLP). 
The number of nodes are $100$ (reservoir layer), $256$ (MLP layer \#1), $192$ (MLP layer \#2), $128$ (MLP layer \#3), and $1$ (output layer). 
Figure \ref{figS:ESN}a illustrates the distribution of indices for each layer. 
The averaged index $\left<\bar{d}_i\right>$ (i.e., node-wise ESP index averaged over nodes in a layer) decreases with an increase in the MLP layer ID [$\left<\bar{d}_i\right>= 0.96$ (reservoir), $1.42$ (\#1), $1.35$ (\#2), $1.24$ (\#3), and $0.287$ (output)].

Next, to double-check whether the ESN is a function of input history or not, we evaluated the conditional Lyapunov exponent (CLE) of the reservoir layer. 
The state equation of the $N$-dimensional ESN is described by 
\begin{align*}
    x_{i,t+1} = \tanh\left(\sum_{j=1}^N W_{ij}x_{j,t} + w_{{\rm in},i} u_t\right), 
\end{align*}
where $x_{i,t}$ is the $i\uth$ state $(i=1,\ldots,N)$, $u_t$ denotes the uniform random input in the range of $[-1,1]$, 
the input weights $w_{{\rm in},i}$ are generated from a uniform random number in the range of $[-0.1,0.1]$, 
and the internal weight matrix $\bW=(W_{ij})$ is also generated from a uniform random number in $[-1,1]$ and then is rescaled such that its spectral radius is equivalent to $\sigma$. 
In this paper, we used $N=100$ and $\sigma=1.3$. 
The Jacobian matrix of ESN is given by 
\begin{align*}
    \bJ_t = \left(\frac{\partial x_{i,t+1}}{\partial x_{j,t}} \right) = \left(\bI - {\rm diag}(\bx_{t+1}\circ \bx_{t+1})\right) \cdot\bW. 
\end{align*}
Using the QR decomposition, we calculate the orthonormal matrix $\bQ_t$ and the upper triangular matrix $\bR_t$ as follows: 
\begin{align*}
    \bJ_t\bQ_{t-1} = \bQ_t \bR_t ~ (t=1,2,\ldots). 
\end{align*}
Note that $\bJ_0 = \bQ_0 \bR_0$. 
Using the diagonal elements of $\bR_t$, we calculate the Lyapunov exponent $\lambda_i$ as follows: 
\begin{align*}
    \lambda_i = \frac{1}{T} \sum_{t=0}^{T-1} \ln |R_{t,ii}| ~ (i=1,\ldots,N). 
\end{align*}
As a result, we obtained the maximum conditonal Lyapunov exponent (MCLE) $\lambda_{\rm max} = 1.5\times10^{-2}$. 
This positive exponent and the averaged node-wise ESP index $\left<\bar{d}_i\right>=1.7$ in the reservoir layer indicate that the ESN does not hold the ESP.

\subsection*{TIPC for Each Layer}
To track the input terms transformed through the layers, we illustrated the first- and second-order capacities in each layer. 
The lower rightmost plot in Fig.~\ref{figS:ESN}b shows the TIPC decomposition of the NARMA10's target output, which is composed of TI first- and second-order capacities. 
To solve the NARMA10 task, the output must include nine components [i.e., TI first-order terms $u_{t-\tau}$ ($\tau=1,2,3,10,11,12$) and second-order ones $u_{t-\tau}u_{t-\tau-9}$ ($\tau=1,2,3$)] with an appropriate ratio~\cite{kubota2021unifying}. 
To make these terms in the output layer, the MLP adjusts their amounts in each layer. 
The reservoir layer holds the TI first-order capacity $C_1(\tau)$. 
The MLP increases the required first-order terms with $\tau=10,11,12$ in layers \#1 and \#2 and then adjusts the ratio of the first-order terms in layer \#3 and the output layer. 
Conversely, the ESN does not have TI second-order terms at all. 
The MLP increases these terms in the layers \#1--3 and adjusts their ratio in the output layer. 
The output layer does not sufficiently hold $u_{t-2}u_{t-11}$ and $u_{t-3}u_{t-12}$, which may deteriorate the NMSE score.

\begin{figure*}[tb]
    \centering
    \includegraphics[scale=0.56]{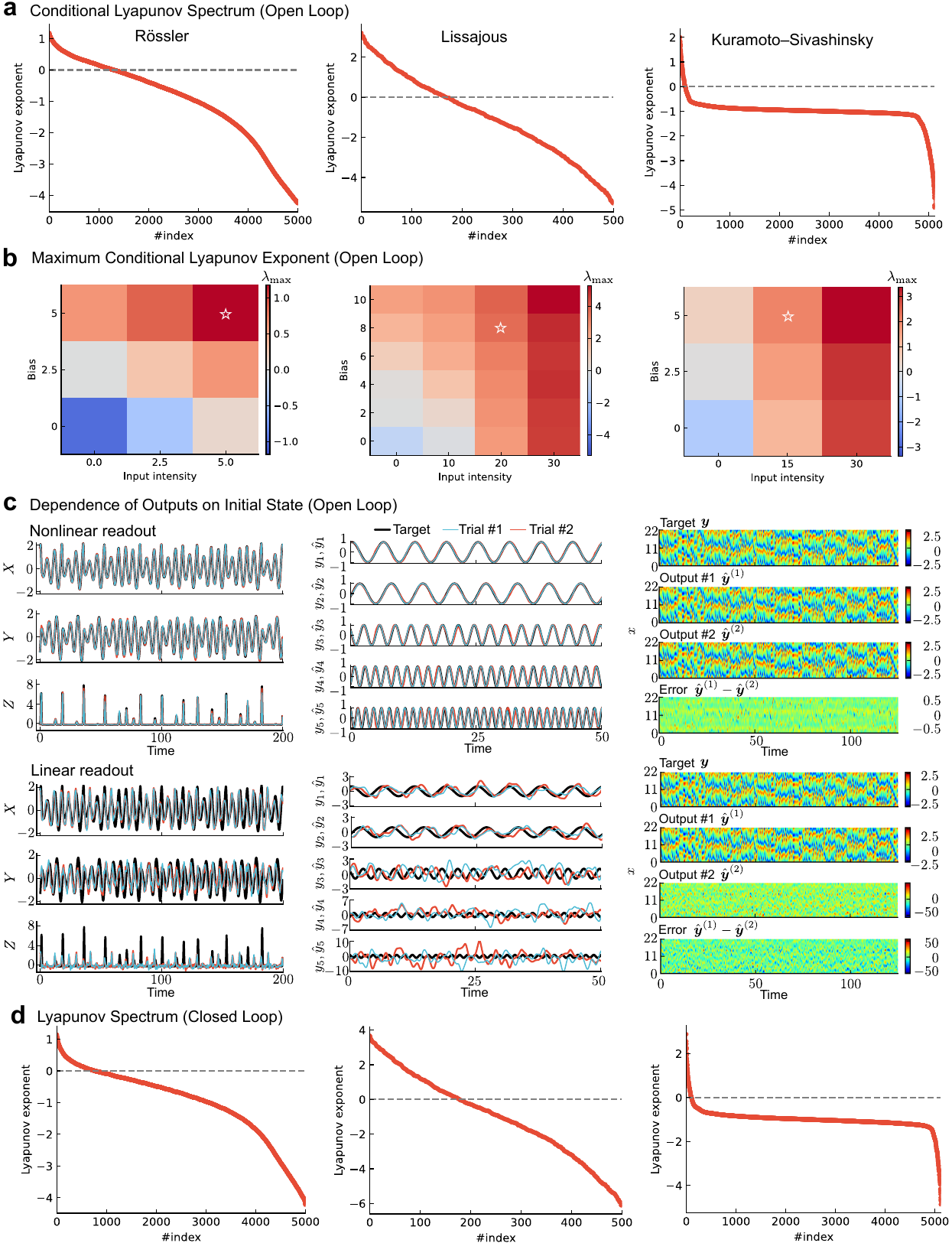}
    \caption{
        \textbf{
        Initial sensitivity of the Lorenz 96 reservoir. 
        } 
        \textbf{a}, Conditional Lyapunov spectra, \textbf{b}, maximum conditional Lyapunov exponents, and \textbf{c}, time series of target and two outputs with different initial states of the open-loop Lorenz 96 reservoir with linear (lower) and nonlinear (upper) readouts, as well as \textbf{d}, Lyapunov spectra of the closed-loop Lorenz 96 reservoir with the target of the R\"ossler model, Lissajous curves, and the Kuramoto--Sivashinsky model. 
        \textbf{a}, The horizontal axis is the index of nodes, and the vertical axis is the conditional Lyapunov exponent in the training phase. 
        \textbf{b}, The horizontal axis is the input intensity $\iota$, the vertical axis is the bias $\mu$, and the color represents the maximum conditional Lyapunov exponent in the training phase. 
        The star marks indicate the parameters used for the embedding tasks. 
        \textbf{c}, In the case of the targets of the R\"ossler model and the Lisssajous curves, the time series of target (black) and two outputs (blue and red) with different initial states of the Lorenz 96 model. 
        In the case of the KS model, the target, two outputs, and error of the outputs, are depicted by colormaps. 
        \textbf{d}, The horizontal axis is the index of nodes, and the vertical axis is the Lyapunov exponent in the test phase. 
    }
    \label{figS:le_lorenz96}
\end{figure*}

\section{Attractor Analysis} \label{sec:attractor_analysis}
\subsection*{Lorenz 96 Model} 
We calculated the CLE to confirm whether the Lorenz 96 reservoir was a function of only input history in the training phase (i.e., the Lorenz 96 reservoir is an open-loop system). 
As shown in Fig.~4 of the main text, we used three types of input: the R\"ossler model, Lissajous curves, and the Kuramoto--Sivashinsky (KS) model. 
The differential equation of the Lorenz 96 model is described by 
\begin{align*}
    \frac{dx_i}{dt} = (x_{i+1}-x_{i-2})x_{i-1}-x_i+\mu+\iota u_i(t)~(i=1,\ldots,N), 
\end{align*}
where $\mu$ and $\iota$ represent the bias and input intensity, respectively. 
We calculated the  conditional Lyapunov spectra using the algorithm described in~\cite{shimada1979numerical}. 
The Jacobian matrix of the Lorenz 96 model is given by 
\begin{align*}
    \bJ = \left(\frac{\partial\dot{x}_i}{\partial x_j}\right) = \begin{pmatrix}
    -1 & x_N & 0 & \cdots & 0 & -x_N & x_2 - x_{N-1} \\
    x_3-x_N & -1 & x_1 & & 0 & 0 & -x_1 \\
    \vdots & & & \ddots & & & \vdots \\
    x_{N-1} & 0 & 0 & \cdots & -x_{N-1} & x_1 - x_{N-2} & -1
    \end{pmatrix}. 
\end{align*}
We define $N$-orthonormal vectors $\delta \bx_i(t)\in\mathbb{R}^N~(i=1,\ldots,N)$ and a matrix $\delta\bX(t) = \left[\delta \bx_1(t)\cdots\delta \bx_N(t)\right]$. 
$\delta\bX(t)$ is updated through the following steps. 
First, we numerically solve the following differential equation using the fourth-order Runge-Kutta method: 
\begin{align*}
    \frac{d(\delta\bX)}{dt} = \bJ \delta\bX 
\end{align*}
to obtain the matrix at the next timestep $\delta\bX(t+\Delta t)$. 
Second, we orthonormalize the matrix using the QR decomposition. The matrix $\delta\bX(t+\Delta t)$ is decomposed into 
\begin{align*}
    \delta\bX(t+\Delta t) &= \bQ(t+\Delta t)\bR(t+\Delta t), 
\end{align*}
where $\bQ$ is the orthonormal matrix and $\bR$ is the upper triangular matrix. 
$\delta\bX$ is replaced by $\bQ$ as follows:
\begin{align*}
    \delta \bX(t+\Delta t) &\leftarrow \bQ(t+\Delta t). 
\end{align*}
Note that we set the matrix at the initial timestep to the identity matrix $\delta\bX(0) = \bI$. 
By repeating this procedure, we obtained $\{\bR(\Delta t),\bR(2\Delta t),\ldots,\bR(M\Delta t)\}$. 
The conditional Lyapunov exponent $\lambda_j$ is calculated with diagonal elements of $\bR(t)$ as follows: 
\begin{align*}
    \lambda_i = \frac{1}{M} \sum_{m=1}^{M} \ln |R_{ii}(m\Delta t)| ~ (i=1,\ldots,N). 
\end{align*}
Figure~\ref{figS:le_lorenz96}a shows the conditional Lyapunov spectra of the Lorenz 96 model in the three embedding tasks. 
The MCLEs of the Lorenz 96 model with the target of R\"{o}ssler model, Lissajous curves, and the KS model were estimated to be $\lambda_{\rm max}=1.18,3.67$, and $2.00$, respectively. 
In all cases, the MCLEs were positive, indicating that the Lorenz 96 systems were not functions of only input history in the training phase. 
The MCLE of the Lorenz 96 model can be negative if we use different parameters, such as $(\mu,\iota)=(0, 0)$, and we selected the parameters with a positive MCLE for the embedding tasks (Fig.~\ref{figS:le_lorenz96}b).

We also calculate a global ESP index to confirm the ESP of output in the open-loop Lorenz 96 reservoir. 
We run the Lorenz 96 reservoir with two different initial states $\bx^{(1)},\bx^{(2)}$ and calculate outputs using both linear and nonlinear readouts to show that the output with nonlinear readout has an ESP. 
Let the $M$-dimensional target and outputs with the two initial states be $\by=(y_1\cdots y_M)^\top$ and $\hat{\by}^{(i)}=\left(\hat{y}_1^{(i)}\cdots\hat{y}_M^{(i)}\right)^\top$ $(i=1,2)$, respectively. 
The global ESP index is defined by 
\begin{align*}
    d = \frac{\sum_{i=1}^M {\rm MSE}\left(y_i^{(1)},y_i^{(2)}\right)}{\sum_{i=1}^M {\rm Var}(y_i)} 
    = \frac{\sum_{i=1}^M \sum_{t=1}^T \left(\hat{y}_{i,t}^{(1)}-\hat{y}_i^{(2)}\right)^2}{\sum_{i=1}^M \sum_{t=1}^T \left(y_{i,t}-\bar{y}_i\right)^2},
\end{align*}
where ${\rm MSE}(\cdot,\cdot)$ and ${\rm Var}(\cdot)$ represent the mean square error between the two time series and the variance of the time series, respectively, and $\bar{y}_i=\sum_{t=1}^T y_{i,t}/T$ is the time average of $y_{i,t}$. 
Figure~\ref{figS:le_lorenz96}c shows two output time series with the target of the R\"ossler model, Lissajous curves, and the KS model, in which the two outputs with nonlinear readout are consistent with the target, but those with linear readout have larger errors than the nonlinear cases.
To quantify these errors, we calculate 10 indices with 10 different initial states of the Lorenz 96 model to average them. 
The averaged global ESP indices (mean $\pm$ standard deviation) with nonlinear readout are $(1.53~\pm~0.01) \times 10^{-2}$ (R\"ossler model), $(5.95~\pm~0.54) \times 10^{-2}$ (Lissjous curves), and $(6.87~\pm~0.13) \times 10^{-2}$ (KS model), while those with linear readout are $0.409~\pm~0.007$ (R\"ossler model), $1.92~\pm~0.19$ (Lissjous curves), and $0.246~\pm~0.003$ (KS model). 
These small indices with nonlinear readouts imply that the TI transformation successfully makes the output function of input history.

Finally, we calculate the MLEs of the Lorenz 96 model in the test phase to verify that the closed-loop Lorenz 96 reservoir is chaotic. 
As shown in Fig.~\ref{figS:le_lorenz96}d, the MLEs of the Lorenz 96 model with the target of R\"ossler model, Lissajous curves, and the KS model are $\lambda_{\rm max}=1.15, 3.50,$ and $2.88$, respectively. 
The positive exponents indicate that the Lorenz 96 reservoir is still chaotic with the feedback signal
instead of the teacher forcing signal.

\begin{figure*}[tb]
    \centering
    \includegraphics[scale=1.0]{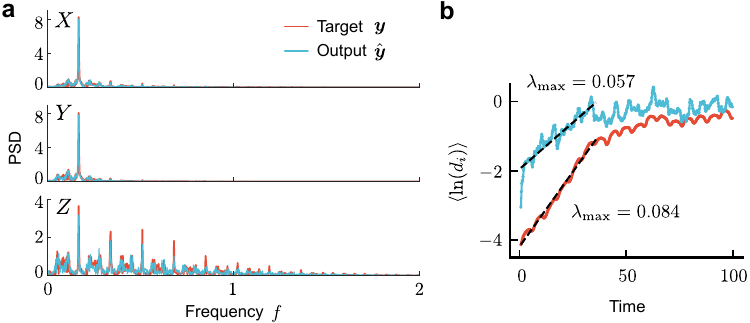}
    \caption{
        \textbf{
        Attractor analysis in the embedding task of the R\"ossler model. 
        } 
        \textbf{a}, Power spectral density of $(X, Y, Z)$ and \textbf{b}, time series of distance between two states beginning from a neighbor point. 
        The target $\by$ and output $\hat{\by}$ are plotted in red and blue, respectively. 
    }
    \label{figS:psd_le_rossler}
\end{figure*}

\subsection*{R\"ossler Model} 
As shown in Fig.~4 of the main text, we perform the three attractor embedding tasks to demonstrate that a system without ESP can be utilized as a computational resource by the nonlinear readout. 
First, we embed the R\"ossler attractor in the Lorenz 96 reservoir. 
To evaluate correspondence between the original and embedded attractors, we calculate their power spectral density (PSD) and maximum Lyapunov exponent.

To estimate the maximum Lyapunov exponent from the time series, we adopt the Rosenstein algorithm~\cite{rosenstein1993practical}. 
First, we find pairs of time $\{t_{i,1},t_{i,2}\}~(i=1,\ldots,M)$ for the nearest neighbors, such as $||\hat{\by}(t_{i,1})-\hat{\by}(t_{i,2})||<\epsilon$. 
We calculate the time evolution of distance between $||\hat{\by}(t_{i,1})-\hat{\by}(t_{i,2})||$ for each pair as follows: 
\begin{align*}
    d_i(k) = ||\bx(t_{i,1}+k\Delta t)-\bx(t_{i,2}+k\Delta t)||_2. 
\end{align*}
We average the logarithm of distance over pairs as follows: 
\begin{align*}
    \left<\ln d_i(k)\right> = \frac{1}{M} \sum_{i=1}^M \ln d_i(k). 
\end{align*}
We draw this time series and approximate its linear part by a linear regression whose slope represents the maximum Lyapunov exponent $\lambda_{\rm max}$.

Figure~\ref{figS:psd_le_rossler}a shows the PSDs of the target and output, which match well. 
The maximum Lyapunov exponents of the output and target are different (the exponent of the output is $\lambda_{\rm max}=0.057$; that of the target is $\lambda_{\rm max}=0.084$). 
Therefore, the Lorenz 96 reservoir did not embed the attractor with the exact same properties as the original one.

\begin{figure*}[tb]
    \centering
    \includegraphics[scale=0.85]{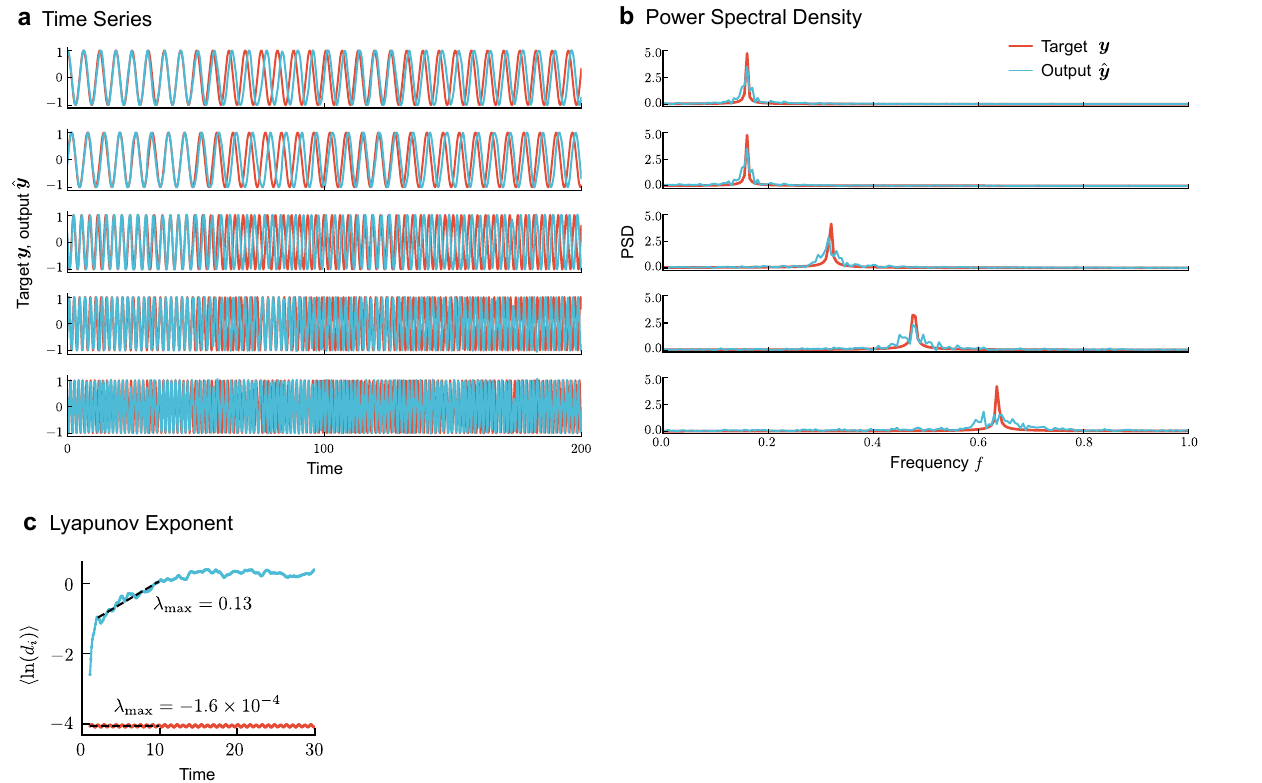}
    \caption{
        \textbf{
        Attractor analysis in the embedding task of the Lissajous curves. 
        } 
        \textbf{a}, Time series, \textbf{b}, power spectral density of the Lissajous curves, and 
        \textbf{c}, time series of distance between two states beginning from a neighbor point. 
        The target $\by$ and output $\hat{\by}$ are plotted in red and blue, respectively. 
    }
    \label{figS:psd_le_lissajous}
\end{figure*}

\subsection*{Lissajous Curves}
Next, we perform the attractor embedding task of the Lissajous curves with the Lorenz 96 reservoir. 
In the main text, comparing the shapes of the embedded attractors with those of the original ones, we show that the Lorenz 96 reservoir can embed the Lissajous curves. 
We also confirm that these attractors are stably embedded by disturbing the feedback signals. 
These embedded attractors do not hold original properties on a long-term basis. 
Figure~\ref{figS:psd_le_lissajous}a shows that the long time series of the target and output do not perfectly match. 
As shown in Fig.~\ref{figS:psd_le_lissajous}b, this difference can be found by the PSDs of output, which are distributed around the peak of the target PSD but have a different distribution. 
The maximum Lyapunov exponent of the target is $0$, but that of the output is positive ($\lambda_{\rm max}=0.13$, Fig.~\ref{figS:psd_le_lissajous}c). 
This outcome is similar to the designed periodic chaos recently reported in~\cite{kabayama2024designing}.

\begin{figure*}[tb]
    \centering
    \includegraphics[scale=1.0]{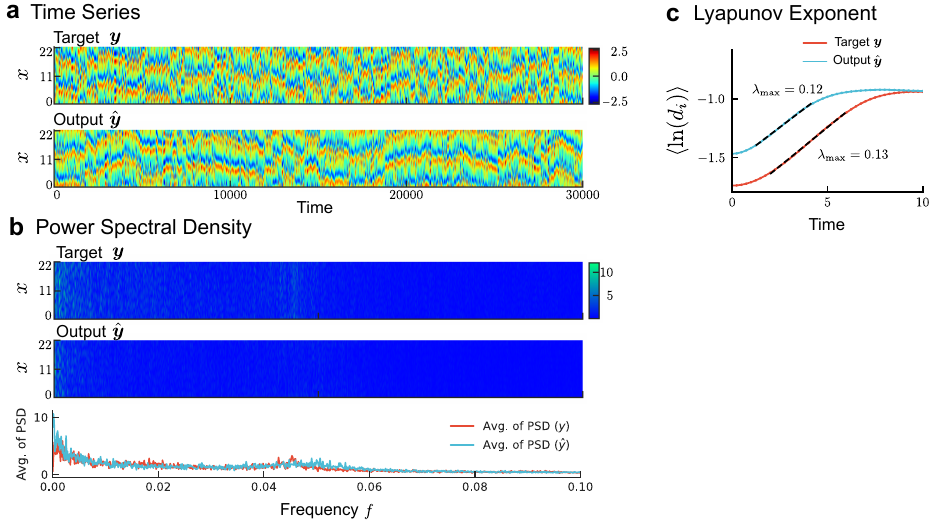}
    \caption{
        \textbf{
        Attractor analysis in the embedding task of the Kuramoto--Sivashinsky model. 
        } 
        \textbf{a}, Time series, \textbf{b}, power spectral densities, and \textbf{c}, time series of distance between two states beginning from a neighbor point. 
    }
    \label{figS:psd_le_kuramoto}
\end{figure*}

\subsection*{Kuramoto--Sivashinsky Model}
Finally, we perform the attractor embedding task of the KS model using the Lorenz 96 reservoir. 
The KS system is defined by the following partial differential equation for the state $y(x,t)$: 
\begin{align}
    \frac{\partial y}{\partial t} + \frac{\partial^2 y}{\partial x^2} + \frac{\partial^4 y}{\partial x^4} + \frac{1}{2} \left( \frac{\partial y}{\partial x} \right)^2 = 0 \label{eqS:kuramoto_sivashinsky}
\end{align}
on a periodic domain $0\le x \le L$ [i.e., $u(x,t)=u(x+L,t)$] with $L=22$. 
We evenly span the space to define the $Q(=64)$ variables 
\begin{align*}
    \by(t) = \left(y(\Delta x, t), y(2\Delta x, t), \ldots, y(Q\Delta x, t) \right)^\top
\end{align*}
with an interval of $\Delta x=L/Q$. 
As shown in Fig.~\ref{figS:psd_le_kuramoto}a, the output $\hat{\by}$ shows a similar spatiotemporal pattern to the target time series. 
To evaluate their correspondence, we calculate their PSDs for each position (Fig.~\ref{figS:psd_le_kuramoto}b). 
Since Eq.~(\ref{eqS:kuramoto_sivashinsky}) is common for the position and the boundary is cyclic, we calculate the PSD averaged over position. 
The averaged PSD of the target and output match well. 
Furthermore, we evaluate the maximum Lyapunov exponents of target $\by$ and output $\hat{\by}$ using the Rosenstein algorithm. 
As shown in Fig.~\ref{figS:psd_le_kuramoto}c, the estimated maximum Lyapunov exponents for the target and output are $\lambda_{\rm max}=0.12$ and $\lambda_{\rm max}=0.13$, respectively, showing good agreement.

\begin{figure*}[tb]
    \centering
    \includegraphics[scale=0.32]{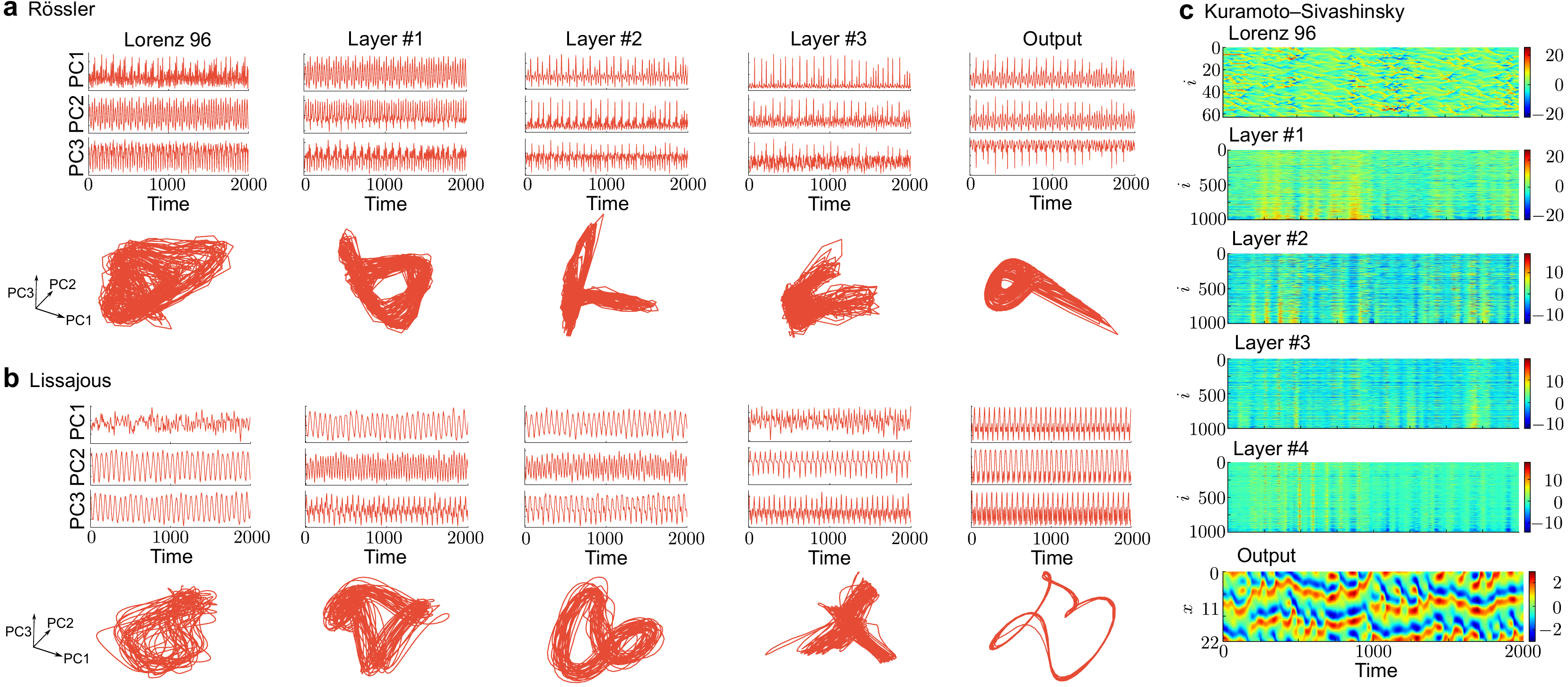}
    \caption{
        \textbf{
        Attractor transition through the MLP layers. 
        } 
        Time series (upper) and three-dimensional plot (lower) of \textbf{a}, the R\"ossler model and \textbf{b}, Lissajous curves. 
        Time series of the \textbf{c}, Kuramoto--Sivashinsky model. 
        \textbf{a}, \textbf{b}, First three principal components $({\rm PC1}, {\rm PC2}, {\rm PC3})$ were plotted. 
        \textbf{c}, Only the first $64$ nodes in the Lorenz 96 reservoir layer were plotted. 
        In layers \#1--4, the nodes were sorted in the order of the two nodes' correlation and only the first 1,000 nodes were plotted. 
    }
    \label{figS:attractor_transition}
\end{figure*}

\section{Attractor Transition Through MLP Layers} 
To visualize the transformation through the MLP layers, we depict a 3D plot or time series for each layer. 
Figure~\ref{figS:attractor_transition}a and \ref{figS:attractor_transition}b illustrate the 3D trajectories for each layer in the attractor embedding task of the R\"ossler model and Lissajous curves, respectively. 
We perform principle component analysis to extract three-dimensional states from each layer and plot first three principal components $({\rm PC1},{\rm PC2},{\rm PC3})$. 
In the case of the embedding task of the KS model, we depict the time series for each layer in Fig.~\ref{figS:attractor_transition}c. 
In the three cases, each layer shows different trajectories or time series, which are linearly combined by the skip connections of the output layer to emulate the target.

\section{Examples of Time-Invariant Transformation} \label{sec:examples_TI_transformation}
We introduce TI transformation for nonstationary and periodic systems.

\subsection*{Nonstationary System with Detrending}
A nonstationary system with input shows TV behavior. 
Here, we adopt the trend stationary model~\cite{nelson1982trends} to show its TI transformation. 
Let the input at the $t\uth$ step be $u_t$, and the state $x_t$ is described by the sum of the function of time $f(t)$ and stationary process $g(u_{t-1},u_{t-2},\ldots)$ as follows: 
\begin{align}
    x_t = f(t) + g(u_{t-1},u_{t-2},\ldots). \label{eqS:trend_stationary}
\end{align}
where $g(u_{t-1},u_{t-2},\ldots)$ is TI and has a fixed time-averaged value. 
The detrending works as a TI transformation. 
For example, if the trend stationary model is represented by 
\begin{align}
    x_t = at + b + c u_{t-1}, \label{eqS:example_trend_stationary}
\end{align}
we can calculate the average of $N$ past states $\bx_t=\{x_{t-N+1},\ldots,x_t\}$ as follows: 
\begin{align*}
    \bar{x}_t &= \frac{1}{N} \sum_{i=0}^{N-1} x_{t-i} = at - \frac{N-1}{2}b + c\bar{u},
\end{align*}
where $\bar{u}=\sum_{i=1}^N u_{t-i}/N$ converges to a fixed value if $N$ is sufficiently large. 
Under this condition, the detrending can form a TI output: 
\begin{align}
    f(\bx_t) = x_t - \bar{x}_t = \frac{N+b}{2} + c(u_{t-1} - \bar{u}). \label{eqS:TI_transformation_trend_stationary}
\end{align}
Note that Eq.~(\ref{eqS:TI_transformation_trend_stationary}) works with the general representation of the stationary process $g(u_{t-1}, u_{t-2}, \ldots)$ instead of $cu_{t-1}$ in Eq.~(\ref{eqS:example_trend_stationary}). 
This transformation can be realized by a readout with memory. 
For example, if we use a dynamical system with linear readout as a readout of generalized reservoir computing, and it has a memory of $N$ past states, the dynamical readout can realize Eq.~(\ref{eqS:TI_transformation_trend_stationary}) by a weighted sum of delayed state series.

The detrending was utilized in ecological reservoir computing as post-processing~\cite{ushio2023computational}. 
In this study, the number of cultured unicellular organisms, {\it Tetrahymena}, was used as a reservoir state and was controlled by temperature input. 
The state increases with time, leading to a nonstationary state, which is transformed into a stationary state by the detrending technique. 
These manual operations remove the time dependency from the system states without ESP, heuristically converting them into states with ESP.

\subsection*{Periodic System with Envelope Extraction
} 
We can transform a periodic system driven by input into a TI output using envelope extraction. 
Let the input at the $t\uth$ step be $u_t$. 
We assume that the state $x(t)$ is given by 
\begin{align*}
    x(t) = a(u_{t-1},u_{t-2},\ldots) \cos(\Omega t), 
\end{align*}
where the amplitude $a(u_{t-1},u_{t-2},\ldots)$ is a function of input, and $\Omega$ is the eigenfrequency. 
Let the Fourier transform of $a(u_{t-1},u_{t-2},\ldots)$ be $A(\omega)$. 
Assuming that $A(\omega)=0$ if $\omega>\Omega$, the Hilbert transform of $x(t)$ gives the following relation~\cite{feldman2011hilbert}: 
\begin{align*}
    H[x(t)] = \hat{x}(t) = a(u_{t-1},u_{t-2},\ldots) \sin(\Omega t). 
\end{align*}
Finally, we can obtain the envelope as the time-invariant transformation 
\begin{align*}
    f[x(t)] = \sqrt{x(t)^2+\hat{x}(t)^2} = a(u_{t-1},u_{t-2},\ldots). 
\end{align*}
In the real STO reservoir, some studies employed the Hilbert transformation for the envelope extraction, which is applied as post-processing after observing the entire time series~\cite{tsunegi2023information}. 
Others used an envelope detector with a diode~\cite{torrejon2017neuromorphic,grollier2020neuromorphic}, which contributes to real-time computation.

\end{document}